\documentclass[twocolumn,english,aps,pre,showpacs,superscriptaddress]{revtex4-2}

\usepackage{graphicx}% Include figure files
\usepackage{dcolumn}% Align table columns on decimal point
\usepackage{bm}% bold math

\usepackage[colorlinks=true, allcolors=blue]{hyperref}
\usepackage[T1]{fontenc}
\usepackage{mathrsfs}
\usepackage{amsmath}
\usepackage{amssymb}
\usepackage{wasysym}
\usepackage{esint}
\usepackage{babel}
\usepackage{xcolor}
\usepackage{dutchcal}
\usepackage[export]{adjustbox}

\usepackage{listings}

\newcommand{\fconv}{\bm{\mathcal{f}}}
\newcommand{\fcon}{\mathcal{f}}
\newcommand{\dconv}{\bm{\mathcal{d}}}
\newcommand{\dcon}{\mathcal{d}}

\makeatletter
\@ifpackageloaded{dutchcal}{}{%
	\renewcommand{\fcon}{\mathsf{f}}         % regular
	\renewcommand{\fconv}{\bm{\mathsf{f}}}   % bold
	\renewcommand{\dcon}{\mathsf{d}}
	\renewcommand{\dconv}{\bm{\mathsf{d}}}
}
\makeatother

\definecolor{mygreen}{rgb}{0,0.6,0}
\definecolor{mygray}{rgb}{0.5,0.5,0.5}
\definecolor{mymauve}{rgb}{0.58,0,0.82}

\begin{document}

\title{Unsupervised machine learning phase classification for the Falicov-Kimball model}

\author{Luk\'a\v{s} Frk}

\address{Department of Condensed Matter Physics, Faculty of Mathematics and Physics, Charles University, Ke Karlovu 5, Praha 2 CZ-121 16, Czech Republic}

\author{Pavel Bal\'a\v{z}}

\address{FZU -- Institute of Physics of the Czech Academy of Sciences, Na Slovance 1999/2, 182 21 Prague 8, Czech Republic}

\author{Elguja Archemashvili}

\address{Department of Condensed Matter Physics, Faculty of Mathematics and Physics, Charles University, Ke Karlovu 5, Praha 2 CZ-121 16, Czech Republic}
\address{FZU -- Institute of Physics of the Czech Academy of Sciences, Na Slovance 1999/2, 182 21 Prague 8, Czech Republic}

\author{Martin \v{Z}onda}

\address{Department of Condensed Matter Physics, Faculty of Mathematics and Physics, Charles University, Ke Karlovu 5, Praha 2 CZ-121 16, Czech Republic}

\begin{abstract}

We apply various unsupervised machine learning methods for phase classification to investigate the finite-temperature phase diagram of the spinless Falicov-Kimball model in two dimensions. Using only particle occupation snapshots from Monte Carlo simulations as input, each technique, including a straightforward classification based on principal component analysis (PCA), successfully identifies the phase boundary between ordered and disordered phases, independent of the type of phase transition. Remarkably, these techniques also distinguish between the weakly localized and Anderson-localized regimes within the disordered phase, accurately identifying their crossover, which is a challenging task for standard methods. Among the machine learning approaches used, PCA based analysis outperforms more complex methods, such as neural network predictors and autoencoders. These results underscore the effectiveness of simple unsupervised techniques in examining phase transitions and electron localization in complex correlated systems.

\end{abstract}
\maketitle

\section{Introduction}

Strongly correlated electron systems are renowned for rich phase diagrams. Due to their complicated nature, it is not uncommon for new phases to be identified in materials or even simple models that have already been investigated for decades. Often, the information about the new phase was already present in the old data, but conventional analysis may have overlooked it. In recent years, a promising solution to this problem has emerged through the application of unsupervised and preferably interpretable machine learning (ML) methods~\cite{carleo2019machine,bedolla2020machine,miles2021correlator}. These techniques can autonomously explore a vast parameter space and identify regions exhibiting distinct features, effectively highlighting potential novel phases. Using well-established physical criteria, researchers can then rigorously analyze the properties of these parameter regions. 

However, despite the tremendous progress in this field, the methods are still in development. The ever-growing repertoire of techniques presents exciting opportunities, but their applicability to specific systems and tasks requires rigorous evaluation under controlled conditions. In this respect, the Falicov-Kimball model (FKM) ~\cite{hubbard1963electron,falicov1969simple} assumes a significant role in the realm of correlated electron systems. The model is accessible to exact methods~\cite{gruber1996thefalicov}, such as dynamical mean-field theory (DMFT) in the infinite-dimensional limit~\cite{freericks2003exact,freericks2006nonequilibrium} and sign-problem-free Monte Carlo techniques in finite dimensions~\cite{motome1999amonte,maska2006thermodynamics,zonda2009phase,zonda2012phase}. 
Simultaneously, it has complicated ground-state~\cite{lemanski2002stripe,cencarikova2011formation} and finite-temperature~\cite{tran2006inhomogeneous,zonda2009phase,zonda2012phase,herrmann2018spreading,yang2024shubnikov,le2024swap} phase diagrams, details of which are still in debate.

FKM is routinely used for the investigation of a multitude of phenomena. For example, it has been applied in its several variations for the description of metal-insulator and valence transitions~\cite{plischke1972coherent,michielsen1994bond,farkavsovsky1995falicov,portengen1996linear,czycholl1999influence,byczuk2005metal,lemanski2017extended,haldar2017real,farkavsovsky2019pressure,nasu2019quasiperiodicity,haldar2019universal}, formation of inhomogeneous charge and spin orderings~\cite{lemberger1992segregation,freericks1999phase,Freericks2002,Lemanski2004,tran2006inhomogeneous,cencarikova2011formation,zonda2012phase,debski2016possibility}, 
ferroelectricity \cite{batista2002electronic,FarkyFero2002,FarkyFero2008,Schneider2008,Golosov2013,farkavsovsky2023hartree,do2024mass}, cold atoms in optical lattices~\cite{MaskaPRL2008,Iskin2009,MaskaPRA2011,HuMaska2015},
transport through layered systems~\cite{Freericks2001,FreericksBook2006,Hale2012,Kaneko2013,zonda2019nonequilibriom}
and other non-equilibrium phenomena~\cite{freericks2006nonequilibrium,EcksteinPRL2008,smorka2020electronic}. 

Crucial is also its role in the development and benchmarking of new methods for strongly correlated systems~\cite{janis1991anew,freericks2003exact,freericks2006nonequilibrium,maska2006thermodynamics,zonda2012phase,kapcia2020extended}. In recent years, these include ML techniques of various kinds. Artificial neural network techniques have been used to boost the equilibrium~\cite{huang2017accelerated,liu2022machine} as well as kinetic~\cite{zhang2022anomalous} Monte Carlo (MC) simulations of FKM. Fully interpretable unsupervised prediction-based methods built on deep learning, as well as physically motivated mean-based strategies, have recently been introduced for the rich ground-state phase diagram of the FKM~\cite{arnold2021interpretable} and have paved the way for more general strategies~\cite{arnold2022replacing}. Recently, Richter-Laskowska et al.~\cite{richter2023learning} applied the Learning by Confusion (LbC) approach to investigate the critical temperature $T_c$ between the ordered and disordered phases of the half filled spinless FKM. The authors have found that $T_c$ is correctly determined by LbC only in the case
of a continuous phase transition that takes place in the regime of medium and strong electron correlations. For a discontinuous phase transition, there is a large ambiguity. Arnold et al. ~\cite{arnold2024mapping} subsequently proposed a possible remedy to this problem, based on a modified indicator of phase transition.

However, as we show in our work, several unsupervised techniques, some of which are simpler than LbC, can locate the correct position of the phase boundary between the ordered and disordered FKM phase regardless of the type of phase transition. The identification of other phase boundaries, which further divide both ordered and disordered regions of the model, proved to be more challenging. 
Nevertheless, we demonstrate that unsupervised techniques based on principal component analysis (PCA)~\cite{wang2016discovering,hu2017discovering} and to the same extent also the prediction-based method and autoencoder-based classification can automatically distinguish the weakly localized bad metal regime and the Anderson-localized insulator regime and locate their crossover. This is important because the problem of electron localization in two-dimensional systems is notoriously difficult to analyze~\cite{markos2006numerical,suntajs2023localization}. Considering the long history of FKM, this crossover was identified only recently. Its existence in finite-sized systems was proven by Antipov et. al~\cite{antipov2016interaction} via investigation of the energy-resolved inverse participation ratio (IPR) and conductivity. Yet, unsupervised ML techniques can pinpoint this crossover directly from the particle occupation snapshots.
Our results underscore the power of straightforward unsupervised machine learning techniques in identifying both first and second order phase transitions, as well as distinguishing various electron localization regimes. Here presented methods are immediately applicable to any model of correlated electrons coupled to classical degrees of freedom (see e.g., ~\cite{liang2013nematic,buhler2000magnetic,yin2010unified,mondal2021when,smorka2022nonequilibrium,hu2015interplay,diaz2021majorana,elbracht2020topological,maska2021unconventional,smorka2025influence}).

The rest of the paper is structured as follows: In Sec.~\ref{sec:FKM} we introduce the spinless FKM at half filling and briefly discuss its finite-temperature phase diagram. Section~\ref{sec:DP} describes the Monte Carlo method used and the details of data production. In Sec.~\ref{sec:MLmethods} we present the machine learning methods employed, including classification methods based on Principal Component Analysis (PCA) in Sec.~\ref{sec:PCA}, the prediction-based classifier with neural networks used as predictors in Sec.~\ref{sec:PBM}, and the autoencoder method in Sec.~\ref{sec:aenc}. The results of the corresponding automatically constructed phase diagram are provided in Sec.~\ref{sec:Results}. The main conclusions are summarized in Sec.~\ref{sec:Concl}. Additional technical details, supporting arguments, alternative PCA analyses, and an investigation of the model beyond half filling are provided in the appendixes.

%--------
\section{Model and data generation \label{sec:Model-and-data}}

\subsection{Falicov-Kimball model \label{sec:FKM}}
The Hamiltonian of the spinless FKM can be written as  
\begin{eqnarray}
H & = & -t\sum_{\left\langle i,j\right\rangle }\left(d_i^{\dagger}d_j^{\phantom{\dagger}}+d_j^{\dagger}d_i^{\phantom{\dagger}}\right)\nonumber \\
 &  & + U\sum_j\left(f_j^{\dagger}f_j^{\phantom{\dagger}}-\frac{1}{2}\right)\left(d_j^{\dagger}d_j^{\phantom{\dagger}}-\frac{1}{2}\right)\label{eq:Model}
\end{eqnarray}
where the first term describes nearest-neighbor hopping of spin-less
fermions on a lattice. Here, $t$ is the hopping integral that we use as the energy unit (i.e., $t\equiv 1$) throughout the paper. The second term represents a repulsive local interaction between the localized (heavy) $f$ particle and the itinerant $d$ (light) particle at the same lattice site. The factors $-\frac{1}{2}$ secure the half filling conditions $N_{f}+N_{d}=L$ for the chemical potential $\mu=0$, with $N_{f(d)}$ being the total number of $f$ ($d$) particles. $L=l^2$ is the total number of lattice points, and $l$ its linear size. We investigate two-dimensional square lattices with periodic boundary conditions. In the main text, we focus on the half filling condition, while additional results for cases far from this point can be found in Appendix~\ref{app:segregated}.

\begin{figure}
	\includegraphics[width=1\columnwidth]{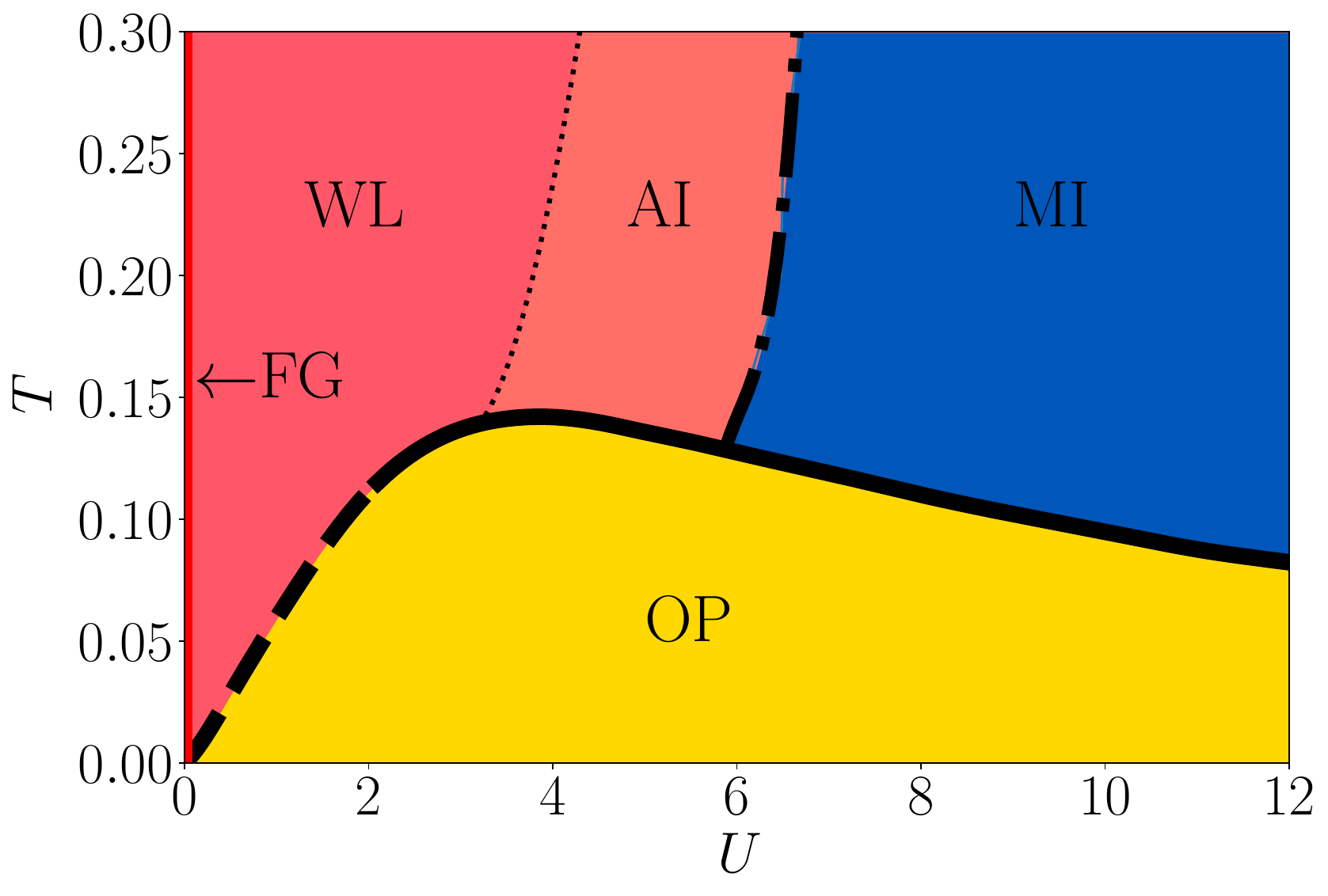}
	\caption{Illustration of the simplified phase diagram of the half filled FKM in two dimensions at finite lattice. The yellow area represents the CDW ordered phase (OP). Other colors signal various types of disordered phases (DP). The phase transition between OP and DP is of the first order at weak interaction $U$ (dashed black line) and of the second order for intermediate and strong $U$ (solid black line). The disordered region is divided (dot-dashed line) between the strong interaction regime (blue), where the system is a Mott insulator (MI), and the weak interaction regime (shades of red). For a finite-sized system, the latter continuously evolves from the Fermi gas (FG) at $U=0$ through the weak localization (WL) region into the Anderson insulator (AI) region at $4\lesssim U \lesssim 6$. Although AI and MI are distinct phases, it is sometimes advantageous to group them together as a disordered insulator (DI) regime. The positions of particular boundaries were taken graphically from Ref.~\cite{antipov2016interaction}.  \label{fig:phd_il}}
\end{figure}

\paragraph*{Phase diagram:} An illustration of the simplified phase diagram of this model is shown in Fig.~\ref{fig:phd_il}. 
It consists of two main regions, namely, the ordered phase (OP) and the disordered one (DP). In OP (yellow), both the $f$ and $d$ particles form a charge density wave (CDW) ordering~\cite{Brandt1989,Brandt1990,Brandt1991,ChenPRB2003,zonda2009phase}.
Although not highlighted in Fig.~\ref{fig:phd_il}, the ordered phase is not homogeneous. It contains gaped and gapless regimes (with respect to the Fermi level)~\cite{Hassan2007,MatveevPRB2008,Lemanski2014,Kapcia2019,zonda2019gapless}, whose boundaries are not yet fully established for finite-dimensional systems.

The critical temperature for CDW ordering in the FKM varies with 
$U$, reaching a maximum around $U\approx 4$. The phase transition is first-order in the weak interaction regime (indicated by the dashed black line) and second-order for intermediate and strong $
U$ values (solid black line)~\cite{maska2006thermodynamics}.

The disordered phase is also divided into the gapless phase for weak interaction $U$  (DPw), and the gaped phase for strong interaction $U$  where the Mott insulator (MI) forms ~\cite{gruber1996thefalicov,freericks2003exact}. However, this is still not a complete picture. In the two-dimensional case, the DPw phase exhibits Anderson localization~\cite{antipov2016interaction}, which destabilizes the previously predicted metallic-like behavior~\cite{maska2006thermodynamics,zonda2009phase}.
In analogy to the Anderson model, i.e., a model with quenched disorder, the DPw phase is expected to be insulating in the thermodynamic limit for any $U>0$. However, due to the exponentially growing correlation length of the Anderson localization with decreasing $U$, there is a sharp crossover from an Anderson-localized insulating (AI) regime at intermediate $U$ to a weakly localized (WL) regime with metallic-like character for \emph{any finite system size}. Moreover, for typical lattice sizes accessible in numerical simulations, the boundary between these regimes appears to be stable with respect to finite-size scaling~\cite{antipov2016interaction}. Finally, a Fermi gas (FG) is formed at $U=0$. Although AI and MI are distinct phases, it is advantageous to group them together in the text as a disordered insulator (DI) regime.  

\subsection{Data production\label{sec:DP}}

Because the $f$-particle number operators $n^f_j$ are good quantum numbers with respect to the Hamiltonian~(\ref{eq:Model}), they can be replaced by their eigenvalues $\fcon_j=1$ (occupied) or $0$ (unoccupied). 
The system can then be sampled over the space of possible $f$-particle configurations $\fconv$ ($f$-configuration for short) via the classical MC procedure~\cite{motome1999amonte,maska2006thermodynamics,zonda2009phase,zonda2012phase}. 
At each MC update, a quantum problem equivalent to a tight-binding model of $d$-particles in a staggered potential defined by $\fconv$ and $U$ is solved via exact diagonalization 
\begin{eqnarray}
H^f_{\mathrm{S}}(\fconv) &=&  \sum_{i,j} h_{ij}(\fconv)d_{i}^{{\dagger}} d_{j} - \frac{U}{2}(N_f -\frac{L}{2}) \\
&=& \sum_\alpha \lambda_\alpha(\fconv) b_\alpha^\dagger b_\alpha  - \frac{U}{2}(N_f -\frac{L}{2})\label{eq:CentralHamiltonianSimpl}
\end{eqnarray}
where $h_{ij}(\fconv)=U(\fcon_i-\frac{1}{2})\delta_{ij}-t_{ij}$ and $t_{ij}$ are the elements of hopping matrix, which are one for nearest neighbors and zero otherwise.
The respective unitary transformation can be written $\bm{\lambda}(\fconv)=\bm{\mathcal{U}}(\fconv) \bm{h}(\fconv)\bm{\mathcal{U}}^\dagger(\fconv)$ where $\bm{\mathcal{U}}(\fconv)$ is the matrix of eigenvectors and $\lambda_\alpha(\fconv)$ the diagonal matrix of eigenvalues of the Hamiltonian $\bm{h}(\fconv)$ for a particular $f$-configuration that are stored in ascending order.  We are using a grand canonical type of update and, therefore, the total number of particles is fixed only on average by the same chemical potential $\mu=0$ for $f$ and $d$-particles.

The investigation of the FKM phase diagram is often restricted to the analysis of the 
$f$-particles and quantities that can be directly calculated from the eigenvalues of the single-particle problem, such as the (global) density of states (DOS). This has the technical advantage of faster data production as the eigenvectors do not need to be evaluated explicitly. This approach is sufficient when investigating order-disorder phase transitions, phases with different $f$-particle orderings, or the opening of the spectral gap in the DOS in the MI phase. However, when addressing the problem of the localization of 
$d$-particles, necessary for investigation of the transition from the weak localization to the Anderson insulator regime, an analysis of the
$d$-subsystem is essential.

We have generated up to $10^4$ equilibrium snapshots of $f$-configurations and related $d$-particle occupations ($d$-configuration for short) 
\begin{equation}
\dcon_i\equiv\left\langle n_{d}^{i}(\fconv)\right\rangle = \sum_{\alpha=1}^L\frac{\mathcal{U}_{i\alpha}(\fconv)\mathcal{U}_{\alpha i}^{\dagger}(\fconv)}{1 + \exp\left[\beta \lambda_\alpha(\fconv)\right]},\label{eq:nd}
\end{equation}
where $i$ is the lattice point index and $\beta=1/T$ in the inverse temperature of the MC simulation,
for each point on a dense grid of parameters in the range illustrated in Fig.~\ref{fig:phd_il}. The configurations have been obtained in $20$ independent MC runs, and each snapshot was separated by a number of complete MC sweeps to avoid trivial statistical correlations. For most of the lattices, we use a parameter grid with steps $\Delta T=0.005$ and $\Delta U=0.25$, refined if needed. We assume that the snapshots of the $f$ and $d$ configurations, together with the parameters $U$ and $T$, at which they have been obtained, are the only information accessible for learning. A subset of the data used for training models is available at Zenodo repository~\cite{zondazenodo}.

For some of the classification methods, we also took advantage of the classical structure factors of the form
\begin{eqnarray}
    S^{\fcon}_{\bm{q}} &=& \frac{2}{L^2}\sum_{j=1}^L\sum_{k=1}^L\left(\fcon_j\fcon_k-\frac{1}{2}\right) e^{-i\bm{q}(\bm{R}_j-\bm{R}_k)},\\
    S^{\dcon}_{\bm{q}} &=& \frac{2}{L^2}\sum_{j=1}^L\sum_{k=1}^L\left(\dcon_j\dcon_k-\frac{1}{2}\right) e^{-i\bm{q}(\bm{R}_j-\bm{R}_k)},
    \label{eq:SF}
\end{eqnarray}
which is the ML language equivalent to the utilization of the polynomial features of the second order.         

%-----
\section{Machine learning methods for automatic classification \label{sec:MLmethods}}
\subsection{Principal component analysis \label{sec:PCA}}

Principal component analysis (PCA) is one of the most widely used techniques in modern data science~\cite{greenacre2022PCA,shlens2014tutorial}. It is most commonly applied to reduce the dimensionality (number of features) of the dataset. This is done by constructing an ordered orthogonal space of principal components (PC). Here, the first PC represents the directions in the original data space that explains the maximal amount of variance, and each subsequent PC represents the maximum variance left. Let us have a dataset with zero mean that is stored in a matrix $\bm{X}$ of dimension $m\times n$. Here, each line stores a new sample, and each column has a particular measured feature. In our case $n=L$ because we store only the lattice point occupancies, i.e., $f$ or $d$-configurations. 
Technically, PCA aligns with the singular value decomposition (SVD) $\bm{X}= \bm{U}\bm{\Sigma}\bm{V}^T$, 
where $\bm{\Sigma}$ is a diagonal square matrix with an ordered set of positive real singular values $\sigma_1\geq\sigma_2\geq\dots\sigma_L$. The columns of $\bm{v}$ (principal components) form a set of orthonormal eigenvectors of the covariance matrix $\bm{X}^T\bm{X}\bm{v}_j = \sigma_j^2\bm{v}_j$ and $\sigma_j^2$ are therefore respective variances.  

The normalized variances, defined as  $\epsilon_j=\sigma^2_j/\sum_{j} \sigma^2_j$ are commonly known as explained variance ratios (EVRs). 
EVRs are crucial for the reduction of dimension, because one can set a threshold value for the total explained variance $\epsilon_T$ needed and keep only first $N_\mathrm{PC}$ principal components such that $\sum_{j=1}^{N_\mathrm{PC}}\epsilon_j\simeq \epsilon_T$. 
However, in this way, PCA can also be used to uncover underlying structures in the data and consequently signal a phase transition~\cite{wang2016discovering,hu2017discovering}, when the parameters change. Basically, a small $N_\mathrm{PC}$ for $\epsilon_T\approx1$ signals a (linearly) ordered data, while a large $N_\mathrm{PC}$ signals a disordered set. 

We apply two versions of the automatic PCA phase classification, which we call the \emph{global} and \emph{local} variants. The \emph{global variant} aligns with the way the method is typically used~\cite{wang2016discovering,hu2017discovering}. The PCA is first applied to all the data (all values of $U$ and $T$). Then the projections of the original configurations at particular parameters to the reduced PC space are investigated. The larger the portion of the variance that falls to the first (first few) PC dimensions, the more ordered the configuration. These results are discussed in the Appendix~\ref{app:globalPCA}.

In the main text, we focus on the \emph{local variant}. Here, PCA is applied to the set of configurations for each combination of parameters $U$ and $T$ independently. In the first step, the analysis of $N_\mathrm{PC}$ can be used to estimate the number of PCs necessary to identify differences in a particular phase. 
We have found that in the case of order-disorder transition, one can simply focus on the first EVR, which can be interpreted as the order parameter, and the second EVR which plays a role similar to charge susceptibility~\cite{wang2016discovering,hu2017discovering}. 
The reason is that the perfectly ordered phase at $T=0$ (and $U\neq 0$) is a simple CDW ordering. Due to translation invariance, PCA rotates such a set completely to just two points lying in the first PC. 
This means that the first EVR will be one and the rest EVRs will be zero. In the opposite limit $T\rightarrow\infty$ of the disordered phase the EVR is evenly distributed among all PC thus $\epsilon_j\approx1/L$ for all $j$. Nevertheless, PCA can be useful even far from half filling, as shown in Appendix~\ref{app:segregated}. 

This PCA variant's locality is both a drawback and an advantage. It requires more data than the global approach, as each phase diagram point is analyzed separately. However, it remains unaffected by data distribution in parameter space and works even with uneven phase representation. This is crucial, as class imbalance may contribute to some failures of the standard LbC~\cite{arnold2024mapping}. Additionally, it enables independent analysis of different parameter space cuts.

A crucial finding of our work is that when the analysis of the $f$ and $d$ configurations is combined, the PCA reveals the sharp crossover between the WL and AI regimes of the FKM.

PCA is a nonparametric method that does not require additional information or tuning beyond the data itself. Its linear nature makes automatic phase analysis simple, fast, and interpretable, but also limits its ability to capture complex non-linear orderings. A partial solution to this limitation is kernel PCA (kPCA), which introduces nonlinearity through the use of kernels. However, these kernels must be carefully designed for each specific problem, requiring prior knowledge of the data structure. This contradicts the main objective of our study. Consequently, we explored other methods that, although more complex, do not suffer from this limitation.

\subsection{Prediction-based method \label{sec:PBM}}

The prediction-based method, initially introduced by Sch\"{a}fer and L\"{o}rch~\cite{schafer2019predmethod}, operates on a straightforward concept. A machine learning (ML) model is trained to infer system parameters from simulated or measured data. Given that the data are expected to be similar within a phase, the predictor may not be able to precisely determine the exact values of the parameters used, e.g., 
$U$ and $T$ for FKM. However, it will indicate parameters that are within the same phase as the target ones. In the idealized case where the configurations are exactly the same within a phase, the inferred points are in the center of the mass of a particular phase ~\cite{schafer2019predmethod}. By analyzing a vector field, where the vectors point from the true (target) parameters to the inferred ones, the locations of phase boundaries can be identified on the basis of where the largest divergence in the field occurs. Technically, the predictor is trained using a mean-square error (MSE) loss function defined as
\begin{equation}\label{eq_mseloss}
\mathcal{L}_{\rm MSE} = \frac{1}{N_{\rm p}N_{\rm k}} 
\sum_{\bm{p}} \sum_{\bm{k}} 
\|\bm{p}-\hat{\bm{p}}\left(\bm{k}\right)\|^2,
\end{equation}
where the sum runs over all sampled points $\bm{p}$ in the parameter space and all $N_{\rm k}$ training configurations $\bm{k}$ at each point $\bm{p}$. Here, $\hat{\bm{p}}(\bm{k})=(\hat{U},\hat{T} )$ denotes the predictions of the predictor given a particular configuration $\bm{k}$. As a rule, predictors have been trained on different sets of configurations $\bm{k}_p$ that were used for the final predictions. We have split the total data into training, validation, and prediction sets, as common in ML.

The final predictions $\bm{\hat{p}}_f$ at parameters $\bm{p}$ are obtained by averaging 
\begin{equation}
\bm{\hat{p}}_f(\bm{p})=\frac{1}{N_{k_p}}\sum_k \bm{\hat{p}}(\bm{k}_p)
\end{equation}
over the prediction set. The vector field is then constructed from the vectors $\bm{\delta p} = \bm{\hat{p}}_f(\bm{p}) - \bm{p}$. Finally, the vector field divergence
\begin{equation}
\nabla_{\boldsymbol{p}} \cdot \boldsymbol{\delta p} = \left.\frac{\partial \delta U}{\partial U}\right|_{\bm{p}} + \left.\frac{\partial \delta T}{\partial T}\right|_{\bm{p}}
\end{equation}
is calculated based on the predictions $\hat{\bm{p}}$ for each sampled point $\bm{p}$ in parameter space. This is done using the symmetric difference quotient
\begin{align}\label{eq_derivatives}
\left.\frac{\partial \delta U}{\partial U}\right|_{\bm{p}} &\approx \frac{\delta U\left(U+\Delta U, N_{f}\right) - \delta U\left(U-\Delta U, N_{f}\right)}{2 \Delta U},\\
\left.\frac{\partial \delta T}{\partial T}\right|_{\bm{p}} &\approx \frac{\delta T\left(U, T+\Delta T\right) - \delta T\left(U, T - \Delta T\right)}{2 \Delta T},\notag
\end{align}
where $\delta U = \hat{U} - U$ and $\delta \rho = \hat{\rho} - \rho$. In the ideal case, a large positive divergence signals the phase boundary.  
\begin{figure}
	\includegraphics[width=0.49\columnwidth,valign=t]{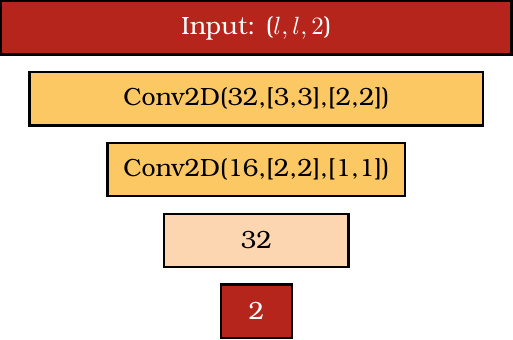}
        \includegraphics[width=0.49\columnwidth,valign=t]{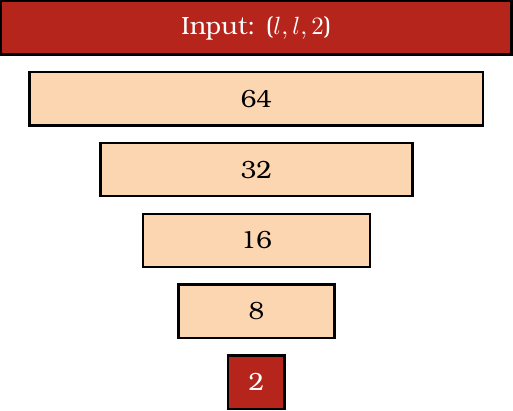}
	\caption{\textit{Left:} CNN predictor with two convolution layers and two dense (fully connected) layers. The first number in a convolution layer is the number of filters, followed by kernel size, and strides. The number in a dense layer gives units, i.e., the dimensionality of the layer output space. \textit{Right:} FCNN predictor with five regular densely-connected layers. The first two layers were followed by batch normalization not shown in the picture for simplicity. In both predictors ReLU activation function was used for all hidden layers and they were set with the Adam optimizer and mean squared error loss function.   \label{fig:predictor}}
\end{figure}
We use three types of predictors, namely a small convolutional neural network (CNN) a fully connected neural network (FCNN) both described in Fig.~\ref{fig:predictor}, and a random forest regressor (RFR) with 100 estimators. 

\subsection{Autoencoder \label{sec:aenc}}

An autoencoder is a type of artificial neural network used for unsupervised learning. It consists of two main parts: an encoder, 
which compresses the input into a lower-dimensional representation,
called the {\em latent space}, and a decoder, which reconstructs the original data from this compressed version. 
The goal is to minimize the reconstruction error across the training dataset. 
This process enforces effective learning of a meaningful representation of the data.
Autoencoders have a range of practical applications. 
The inputs encoded into the latent space can be used to reveal underlying patterns and key characteristics of the input data. 
In the case of the Ising model, exploration of latent space has been used to distinguish between low and high symmetry phases~\cite{hu2017discovering}.
Additionally, the reconstruction error provides valuable insights: a high reconstruction error may indicate that the input is significantly different from what the model has learned, making autoencoders useful for detecting outliers, anomalies, or changes in the data 
distribution~\cite{sakurada2014anomaly,Kong2023unsupervised}.

Unlike PCA, where principal components have clear mathematical interpretations, the latent variables in an autoencoder are learned unsupervised, without explicit constraints enforcing physical interpretability. As a result, the autoencoder may capture spurious correlations instead of the correct order parameter. However, while autoencoders can still reconstruct inputs even when linear assumptions break down, they may fail to structure the latent space in a way that reflects physically meaningful variables. Therefore, in this study, we prioritize analyzing the reconstruction error of decoded inputs to identify distinct phases.

\begin{figure}
    \centering
    \includegraphics[width=.5\columnwidth]{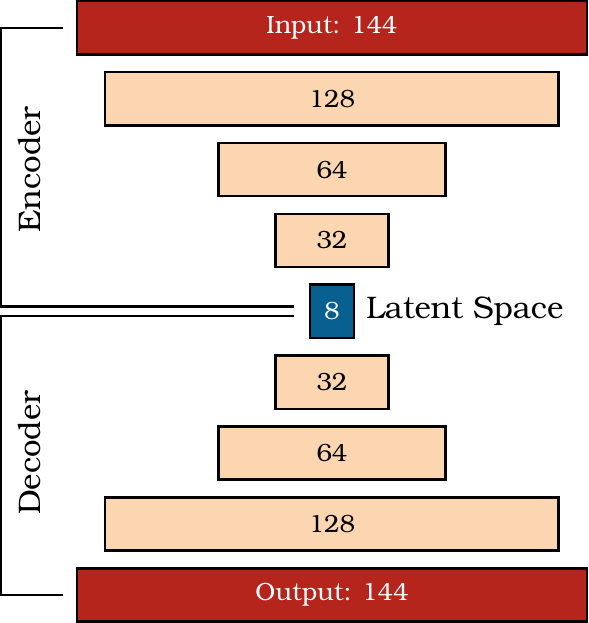}
    \caption{Scheme of an autoencoder used for the analysis of single electron channels configuration. Each block in the scheme consists of an activation-free dense layer of neurons, a batch normalization layer, and an activation layer with ReLU activation function. The output layer features a Sigmoid activation layer. The numbers present the number of neurons in the dense layer of a given block.}
    \label{fig:aenc}
\end{figure}
Here, we use two simple autoencoders of similar structure, as shown in Fig.~\ref{fig:aenc}, to investigate the $f$ and $d$ electron channels separately. Each autoencoder consists of an encoder and a decoder, designed to map the input data to a lower-dimensional latent representation and reconstruct it back.
The input layer has $144$ neurons, representing a batch of configurations for $f$ or $d$ particles. The encoder contains three blocks. Each block contains an activation-free dense layer followed by batch normalization and ReLU activation to introduce nonlinearity. The number of neurons in the dense layers are $128$, $64$, and $32$, respectively.
The latent space, capturing the most essential features of the input data, is represented by a block with 8 neurons.
The decoder is symmetric to the encoder, using three blocks with $32$, $64$, and $128$ neurons, respectively. The final output block of the decoder has $144$ neurons, reconstructing the original input data. 
To keep the values of the output components in the range between $0$ and $1$
the last activation layer restricts the values by a sigmoid activation function.

The difference between the autoencoders processing $f$ and $d$ particle configurations is in the definition of their loss functions, which express the accuracy of the input data reconstruction. Since the configuration vectors of the $f$-channel consist of $0$s and $1$s, a natural choice for the loss function is the binary cross-entropy. 
\begin{equation}
    \begin{split}
        &\mathcal{L}_f(\fconv, \hat{\fconv}) = \\
        &-\frac{1}{L} \sum_{i=1}^{L} 
        \left[ \fcon_i \cdot \log(\hat{\fcon}_i) + (1 - \fcon_i) \cdot \log(1 - \hat{\fcon}_i) \right]\,,
    \end{split}
\end{equation}
where $\fcon_i$ are the targets while $\hat{\fcon_i}$ are the values reconstructed
by the decoder.
On the other hand, for the $d$-channel configurations, containing the itinerant electron densities ranging between $0$ and $1$, we use the loss function in the form of the mean squared error. 
\begin{equation}
    \mathcal{L}_d(\dconv, \hat{\dconv}) = \frac{1}{L} \sum_{i=1}^{L} (\dcon_i - \hat{\dcon}_i)^2\,,
\end{equation}
where $\dcon_i$ and $\hat{\dcon_i}$ are the target and reconstructed electron densities of the $d$-channel, respectively.

\section{Results \label{sec:Results}}

\subsection{PCA phase classification \label{sec:ResPCA}}

We begin with an analysis of the order-disorder transition and demonstrate that PCA of both $d$- and $f$-configurations separately reveals the phase boundary between the CDW and disordered phases automatically. In Appendix~\ref{app:localPCA} we show the number $N_\mathrm{PC}$ of principal components needed to encode 95, 90, 80, and 60\% of the local data variance in the $f$ and $d$ subsystems, thereby clearly demonstrating that there are two distinct regimes. An ordered one where $N_\mathrm{PC}\simeq 1$ and a disordered one where $N_\mathrm{PC}\gg 1$. 
%...
\begin{figure}[ht]
	\includegraphics[width=1\columnwidth]{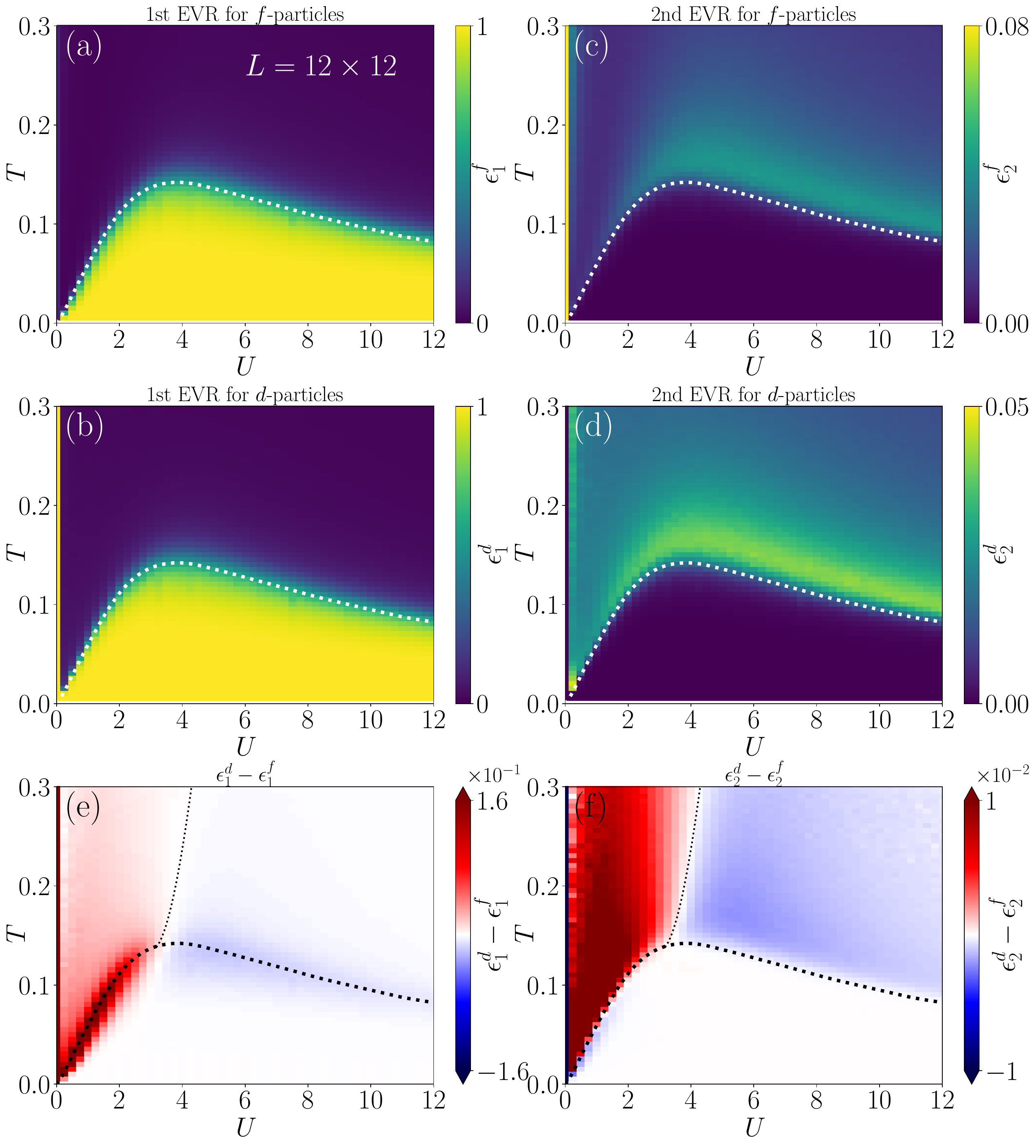}
	\caption{ Results of the local PCA for $L=12\times 12$. Panels (a) and (b) show the value of the first EVR for the $f$ (a) and $d$ (b) particles. Panels (c) and (d) show values of the second EVR. Panels (e) and (f) show the differences in the first (e) and second (f) EVRs for $f$ and $d$ particles highlighting the transition from the WL regime to the DI regime. Dashed lines signal the phase or regime boundaries taken graphically from ref.~\cite{antipov2016interaction}.   \label{fig:evr_map}}
\end{figure}
%...

\begin{figure*}[t]
    \includegraphics[width=2\columnwidth]{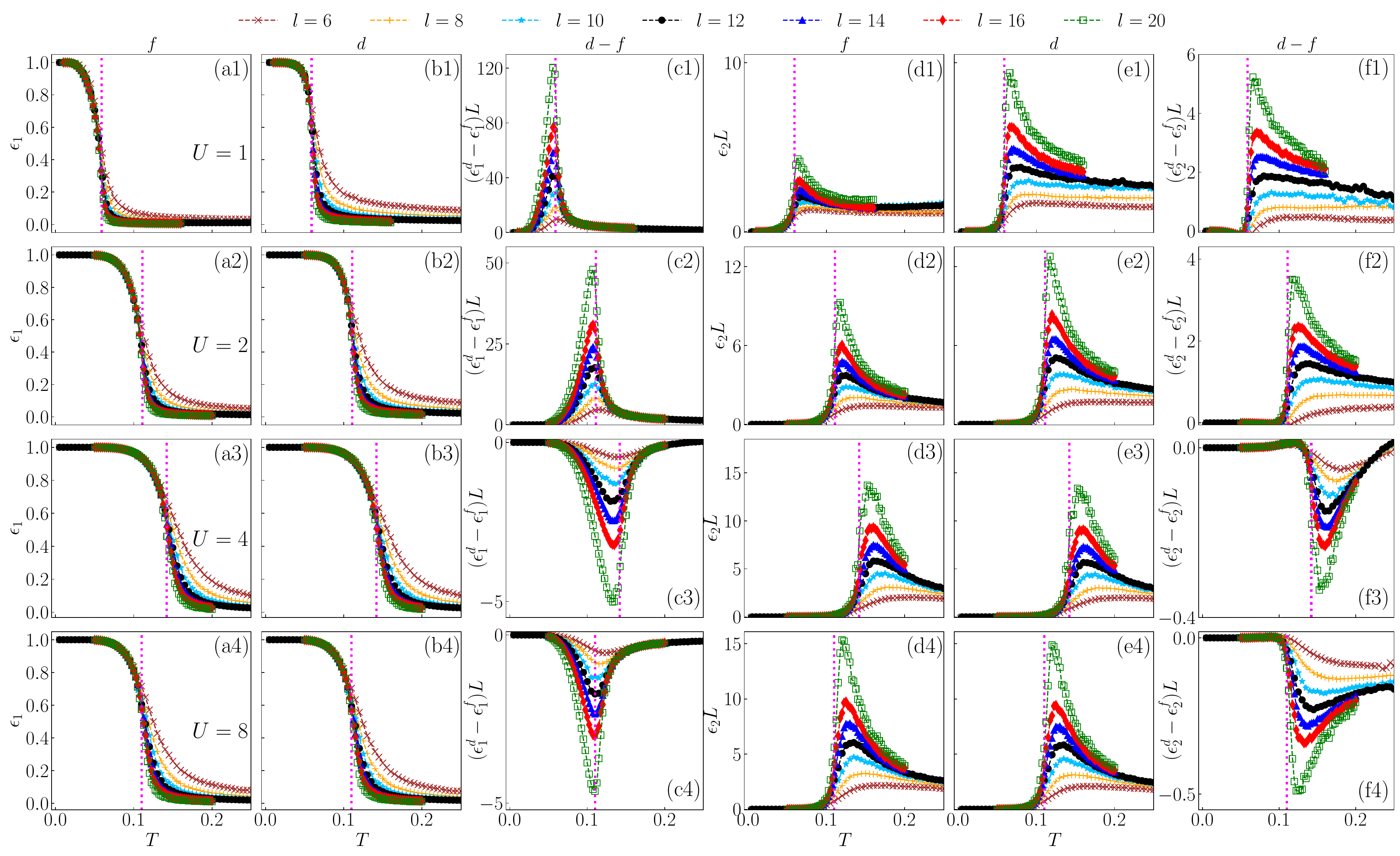}
    \caption{Temperature dependence of various quantities calculated by the local PCA phase classification technique for $U=1$ (row $1$), $2$ (row $2$), $4$ (row $3$) and $8$ (row $4$) and different lattice sizes $L=l\times l$. The columns show: (a) first EVR for $f$ configurations $\epsilon_1^f$, (b) first EVR for $d$ configurations $\epsilon_1^d$, (c) scaled difference $(\epsilon_1^d- \epsilon_1^f)L$, (d) scaled second EVR for $f$ configurations $\epsilon_2^fL$, (e) scaled second EVR for $d$ configurations $\epsilon_2^dL$, (f) scaled difference $(\epsilon_2^d- \epsilon_2^f)L$. Panels (a1)-(b4) show that $\epsilon_1^f$ and $\epsilon_1^d$ play the role of an order parameter. 
    Panels (d1)-(e4) show that $\epsilon_1^f L$ and $\epsilon_2^d L$ have similar properties as charge susceptibility. Panels (c1)-(c4) and (f1)-(f4) show that there is a qualitative difference between weak- and strong-coupling regimes captured by PCA.  
\label{fig:cuts}}
\end{figure*}

Figure~\ref{fig:evr_map} shows the maps of the first [Figs.~\ref{fig:evr_map}(a) and ~\ref{fig:evr_map}(b)] and second [Figs.~\ref{fig:evr_map}(c) and ~\ref{fig:evr_map}(d)] EVRs for the $f$ and $d$ configurations, illustrating their ability to identify the position of the phase boundary. The value of the first EVR approaches one in the ordered phase and drops to approximately $1/L$ in the disordered phase, acting as an automatically constructed order parameter. The transition point coincides with the white dashed line, taken graphically from Ref.~\cite{antipov2016interaction}, which marks the critical temperature. 
The second EVR shows a ridge just above the phase boundary.

Figure~\ref{fig:cuts}, presents finite-size scaling of the first and second EVRs as functions of temperature for different values of $U$. The scaling of the first EVR in panels (ax) and (bx) confirms that $\epsilon_1$ can be interpreted as the order parameter and that the transition becomes sharper with increasing lattice size. The scaling of $L\epsilon_2$ shows that this quantity plays a similar role as the susceptibility, although the phase transition point (vertical dashed line) is better aligned with the inflection point on the left side of the peak than with the moving position of the maximum.
 
The existence of a peak in $L\epsilon_2$ just above the phase transition point can be explained by the growing correlation length between the particles as the temperature decreases toward the critical one. Consequently, a lower number of principal components (but larger than one) is needed to capture most of the variance. The PCA results in panels Figs.~\ref{fig:cuts} (d1)-(d4) and ~\ref{fig:cuts}(e1)-(e4) show that this transitional feature, related to the annealed nature of the $f$ and consequently the $d$ configuration distributions, and its drop in the ordered phase, is rather sharp even for the finite lattices investigated here. Therefore, both the first and second EVRs are good indicators of the order-disorder transition in the whole range of $U$. That is, both the regime of the first-order and the second-order phase transition are captured by the method correctly. An example of how EVR can be used to identify the order-disorder transition even far from half filling is discussed in the Appendix~\ref{app:segregated}.        

However, there are clear differences in the evolution of the EVR between the weak ($U\lesssim 3.5$) and intermediate, respectively, strong interaction regimes, as well as between the EVRs for the configurations $d$ and $f$. Moreover these differences become more pronounced with increasing lattice size if properly scaled, as shown in Figs.~\ref{fig:cuts} (c1)-(c4) and ~\ref{fig:cuts}(f1)-(f4). 
In particular, the transition of the first EVR from one to $\approx 0$ for $d$-particles is sharper than for $f$ particles in the weak interaction regime, i.e., where the WL regime is expected in the disordered phase but broader in the intermediate and strong interacting regime. This manifests itself in sharp positive peaks in $(\epsilon^d_1 - \epsilon^f_1)L$ in Figs.~\ref{fig:cuts} (c1) and ~\ref{fig:cuts}(c2) where $U=1,2$ and sharp negative peaks in panels ~\ref{fig:cuts}(c3) and ~\ref{fig:cuts}(c4) where $U=4,8$. The positions of the peak maxima are just below the critical temperature. When plotted as a map for $L=12\times12$ in Fig.~\ref{fig:evr_map} (e) this highlights the difference between these regimes. 

\begin{figure}[ht]
	\includegraphics[width=1\columnwidth]{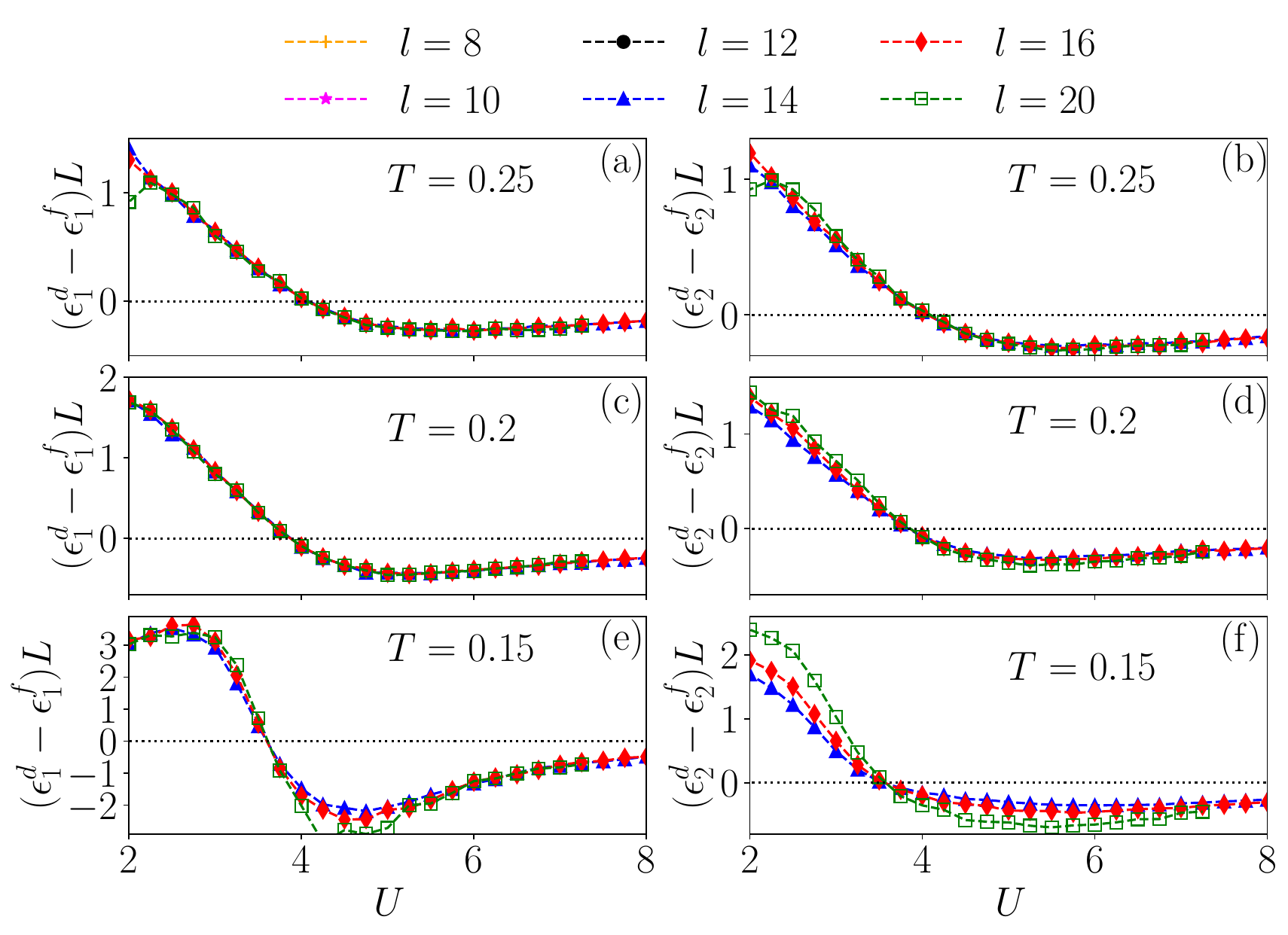}
	\caption{Finite-size scaling of the first and second EVR differences obtained by the local PCA in the vicinity of the expected WL-AI transition. The panels show the evolution of the differences on $U$ for different temperatures larger than the critical one for the order-disorder transition. \label{fig:PCA_Ucut}}
\end{figure}

However, in this sense, a better indicator is the second EVR. 
Figure~\ref{fig:evr_map}(f) shows a map of $(\epsilon^d_2-\epsilon^f_2)L$. This difference is practically zero in the ordered phase (white color), but is positive (red) in the weak interaction regime of the disordered phase and negative (blue) otherwise. Again, the difference at moderate temperatures becomes more pronounced with increasing lattice size, as shown in Figs.~\ref{fig:cuts}(f1)-(f4). Intriguingly, the transition from positive to negative values (white stripe) coincides with the WL-AI transition identified by Antipov et al.~\cite{antipov2016interaction} (dotted line). The position of the boundary is stable with respect to a change in the numerically accessible lattice sizes (Fig.~\ref{fig:PCA_Ucut}).

We interpret the positive value of the difference in the weak coupling regime as a consequence of different characteristics of the $f$ and $d$ configurations. Let us take an example of a very small $U$ and a sufficiently large temperature. In such a case, the EVR of the principal components for $f$-configurations will be close to $1/L$ due to the fact that we are analyzing almost uniformly distributed ones and zeros. In other words, the $f$ configurations are not correlated. On the other hand, the $d$ configurations are almost homogeneous and therefore quite similar to each other, i.e., seem to be correlated. In the limiting case of $U=0$, all $d$ configurations are identical and have lattice point occupancy $0.5$. The variance is zero, which is equivalent to a situation where all the variance is explained by a single component. The FG phase is signaled by $\epsilon^d_1=1$, $\epsilon^d_{n>1}=0$ and $\epsilon^f_i\ll 1$. Extrapolating this to a disordered phase with small finite $U$ the $d$-configurations appear to be more ordered to the PCA simply because they are more homogeneous. The difference decreases as the $d$-configurations become more localized and more sensitive to the position of the $f$-particles at stronger $U$.

\begin{figure}[ht]
	\includegraphics[width=1\columnwidth]{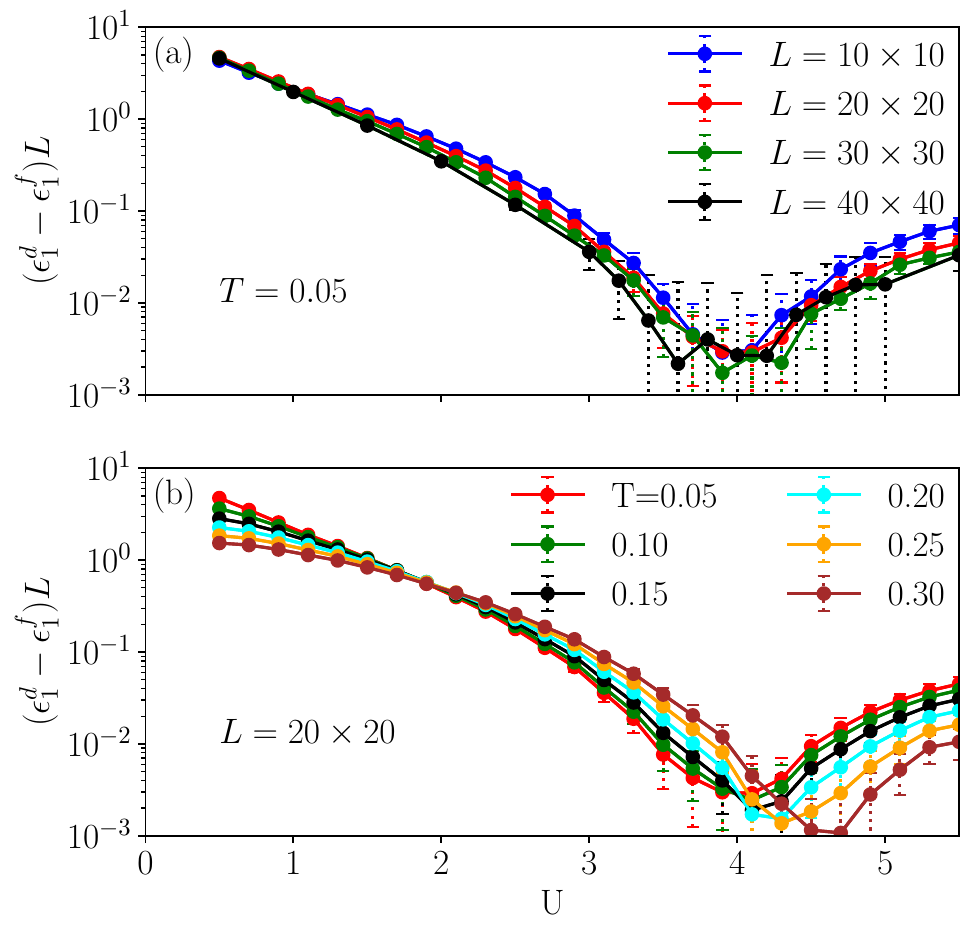}
	\caption{Results of the local PCA for simplified FKM where for each assumed configuration exactly $L/2$ $f$-particles were placed randomly on the grid without taking the $d$ particles into account. This is equivalent to Anderson impurity model. To estimate the error bars the PCA was run independently for $25$-$50$ sets in which PC have been calculated from $10^2$-$10^4$ random configurations.~\label{fig:AIM}}
\end{figure}

Although the negative values of the difference in the strong-coupling regime are more difficult to explain, they are clearly related to the annealed nature of the "disorder", i.e., the distribution of the $f$-particles being dependent on the $d$ particles. We have tested this by analyzing distributions where $L/2$ of $f$ particles are randomly placed on the grid, that is, independently of $d$ particles. The FKM is for such a choice equivalent to the Anderson impurity model. As shown in Fig.~\ref{fig:AIM}, for this case, the difference between the $d$ and $f$ EVRs decreases exponentially with increasing $U$ to the critical $U_c$ and then increases again to some small but, contrary to FKM, positive values. 

The negative values for the FKM are therefore most probably related to the partial checkerboard orderings, which are even above the temperature of the order-disorder transition, still the configurations with the highest probabilities. Consequently, the $f$-configurations appear more ordered to the PCA than the $d$-configurations. Although these are also well localized in the strong-coupling regime, the competition between the checkerboard patterns and homogeneous ones makes the variance more spread among the principal components.
As a consequence, a phase diagram that highlights not only the order-disorder phase transition, but also the finite-size boundary between the WL and DI regimes can be constructed simply by plotting $\epsilon^d_2-\epsilon^f_2$ as shown in Fig.~\ref{fig:evr_map}(f).

The WL-AI boundary appears to be stable with respect to the finite-size scaling for FKM (Fig.~\ref{fig:PCA_Ucut}) and the simplified model (Fig.~\ref{fig:AIM}). However, this should be treated with caution. The localization length of the Anderson localization in the two-dimensional Anderson model shows an exponential divergence $\xi=b\exp(a/U)$ where $a$ and $b$ are some constants~\cite{suntajs2023localization}. Therefore, it rapidly exceeds the linear size of any finite system, even for a macroscopically large one, if $U$ is small enough. Consequently, conclusions about the thermodynamic limit cannot simply be extrapolated from finite grids. Even when the finite-size scaling appears to be compatible with the weak localization in available lattices, it might not survive up to the thermodynamic limit. A similar conclusion also holds for FKM~\cite{antipov2016interaction}. Nevertheless, this also means that at the finite lattices the WL-AI are distinct regimes that can be identified automatically with PCA.        

The fact that a simple PCA approach can identify differences between WL and AI just from the occupation snapshots seems puzzling. More standard approaches require investigating properties such as inverse participation ratio, entanglement, participation entropy, or thorough spectral analysis, just to name a few. However, in a closer analysis, the local PCA contains building blocks similar to some of these methods. For example, the $d$ configurations are observables calculated from single-particle energy eigenstates, and their subsequent EVR analysis resembles both the inverse participation analysis (see, e.g., Eq.~(6) in Ref.~\cite{zonda2019gapless}) and the analysis of the ratio in variances of the matrix elements of local observables in single-particle energy eigenstates (see, e.g., Sec. II.B in Ref.~\cite{suntajs2023localization}). 
This aligns with the recent study by Vanoni and Vitale~\cite{vanoni2024analysis}, which established an analytical connection between the eigendecomposition of the sample covariance matrix and participation entropy. This connection has already enabled the use of this technique to investigate localization in the Anderson model on random regular graphs~\cite{vanoni2024analysis} and various one-dimensional systems~\cite{muzzi2024principal}. While the specific details may differ, the key insight remains: the necessary information is inherently present in the particle snapshots. The next step is to demonstrate that this information can also be effectively extracted using other unsupervised machine learning techniques.

%-----------------------------------------
\subsection{Prediction-based classification}

\begin{figure}[ht]
	\includegraphics[width=1\columnwidth]{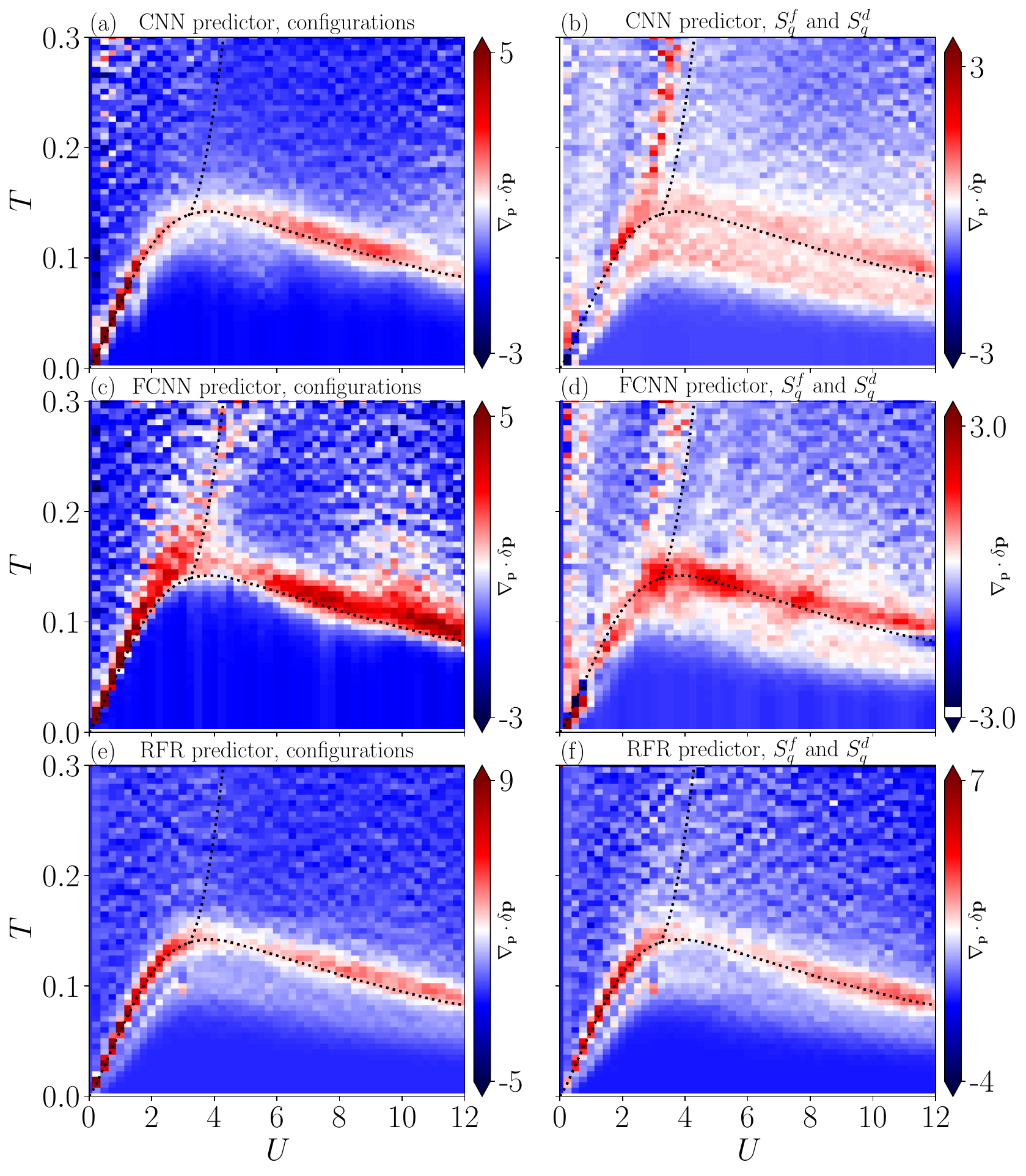}
	\caption{Vector field divergence~\eqref{eq_derivatives} for three types of predictors (rows): (a),(b) CNN predictor; (c),(d) FCNN predictor, and (e),(f) RFR predictor, applied directly to the configurations (first column) and to the classical structure factors~\eqref{eq:SF}. Dashed lines indicate phase transition boundaries, which were taken from ref.~\cite{antipov2016interaction}.   \label{fig:vfd}}
\end{figure}

We use three types of predictors, namely CNN, FCNN, and RFR explained in Sec.~\ref{sec:PBM}. To mitigate some boundary problems, the training data also included results for temperatures higher than $T=0.3$ (up to $T=1$), calculated on a sparser temperate grid than for $T\leq 0.3$. The predictors were first trained to infer $U$ and $T$ from a single pair of $f$ and $d$ configurations for $L=12\times 12$ and then, independently, from their structure factors. The resulting divergences maps of the predictions are shown in Fig.~\ref{fig:vfd}. The panels in the first column depict the results trained directly on configurations with CNN [Fig. ~\ref{fig:vfd}(a)], FCNN [Fig. ~\ref{fig:vfd}(c)]  and RFR [Fig. ~\ref{fig:vfd}(e)] predictors. All clearly highlight the order-disorder phase boundary. The boundary is not as sharp as in PCA; however, this shortcoming of prediction-based classification is known to improve with increasing lattice size~\cite{arnold2021interpretable}. What is more important here is that all predictors work well enough in both weakly and strongly correlated regimes, i.e., capture correctly the transition in both first- and second-order phase transition regimes.

Identifying the transition from the WL and DI regimes is more challenging. CNN [Fig. ~\ref{fig:vfd}(a)] and RFR [Fig. ~\ref{fig:vfd}(e)] predictors applied directly to the configurations do not predict this boundary. We have found that this problem cannot be easily solved by small improvements in these predictors. We have also tested using several (up to 10) randomly chosen configurations (for the same parameters) for training and prediction to infer $U$ and $T$. In theory, this should give the predictor more information about the data distributions. These are expected to be different for the $d$ configurations in the WL and DI regimes. However, this has not solved the problem and has led to a much slower learning process. 

Nevertheless, the FCNN results indicate that this predictor can, in principle, recognize the differences between the WL and DI regimes. This was true even in a case where CNN and FCNN had a comparable number of parameters. The crucial distinction is that small two-dimensional filters of the CNN are not able to capture the long-range particle correlation needed for the WL-DI distinction. For CNN and RFR (and partly also FCNN) predictors, the difference between WL and DI regimes is overshadowed by the clear difference between ordered and disordered phases.      

What worked for the CNN predictor was to use classical structure factors from Eq.~\eqref{eq:SF} instead of configurations as illustrated in panel (b). We stress that the SFs are classical because even for $d$ particles, only their mean values (snapshot measurements) are used instead of calculating the true quantum correlations. 
This means that no additional measurements are needed and that this is just a pre-processing equivalent to using polynomial functions of the second order in the learning. Despite the fact that the same CNN network was trained as a predictor, the method now clearly highlights a rim that divides the disordered phase into two regimes. This boundary is close to, but not perfectly aligned with, the expected WL-DI transition boundary. Nevertheless, it again shows that the information about the WL-DI transition is already present in the $f$ and $d$-configuration.  With data preprocessing in the form of a structure factor, the simple CNN predictor is already sufficient to highlight that there are different regimes in the disordered phase.   

On the other hand, the estimate of the order-disorder phase boundary is worse for the preprocessed data in panel (b) than for the direct approach in panel (a). 
This is despite the fact that this boundary can be predicted from a single data point at $S^{\fcon,\dcon}_{\bm{\pi,\pi}}$ as this is the order parameter for the CDW ordering. The CNN predictor is more sensitive to fluctuations in the structure factors in the broad vicinity of the phase boundary for $U$ from the DI regime, but does not correctly capture the weak coupling one.

Nevertheless, as shown in Figs. ~\ref{fig:vfd}(c) and ~\ref{fig:vfd}(d), these properties are predictor-dependent. For example, the FCNN can identify WL-DI regimes transition, it is sensitive to the difference between the FL and the WL and the order-disorder transition is not as smooth as in Fig. ~\ref{fig:vfd}(c). The RFR predictor gives similar results in the direct approach [Fig. ~\ref{fig:vfd}(e)] and indirect approach [Fig. ~\ref{fig:vfd}(f)].   

The results in Fig.~\ref{fig:vfd} suggest that a more expressive predictor, or even better, physically motivated and tailored predictors~\cite{arnold2022replacing} for the detection of investigated phase transitions could lead to even better results. However, the results presented are already sufficient to show that the prediction-based method can automatically distinguish OP, WL and DI regimes.      

%-----------------------------------------
\subsection{Autoencoder}

Both autoencoders described in Sec.~\ref{sec:aenc} were trained using approximately $10^6$ samples. The training dataset~\cite{zondazenodo} was constructed to uniformly cover the entire range of parameters $U$ and $T$, ensuring a homogeneous representation. Special care was taken to ensure that the dataset samples were uncorrelated. To prevent overfitting, we supplemented the training set with an additional $\approx\! 50 \times 10^3$ samples, obtained at higher temperatures up to $T=1.0$. 

\begin{figure}[htp!]
    \includegraphics[width=1\columnwidth]{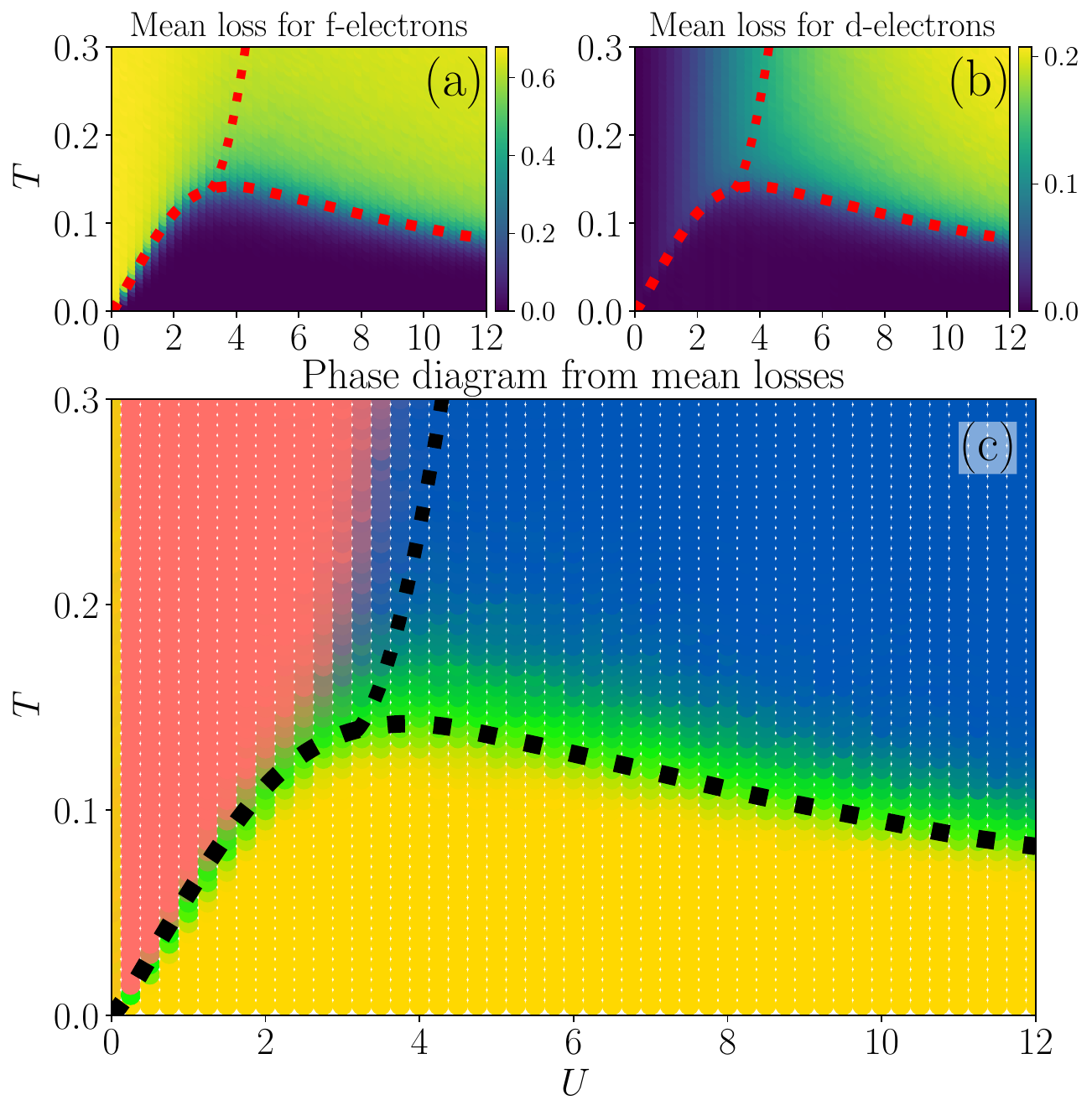}
    \caption{Mean values of Loss function calculated using the validation samples for the (a) $f$-particles, (b) $d$-particles. Plot (c) shows the phase diagram obtained using k-means clustering performed on the loss means.  \label{fig:aenc_loss}}
\end{figure}

After training, model evaluation was performed using another $\approx\! 480 \times 10^3$ validation samples that spanned the entire parameter range. To obtain robust estimates, we averaged the values of the loss function for each parameter pair $(U, T)$. Figure~\ref{fig:aenc_loss}(a) presents the mean loss value of the autoencoder for the $f$-electron channels. It is evident that the loss function is lower in the region covered by the ordered phase configurations. Conversely, the loss function increases as the system enters the disordered phase. This behavior is attributed to the autoencoder's ability to easily learn the chessboard patterns of the ordered phase, while it struggles to reconstruct the highly disordered configurations. In the disordered phase, the trained autoencoder attempts to reconstruct the data using learned ordered-phase patterns, which results in a higher reconstruction error. Consequently, the autoencoder's loss correlates with the boundary between the ordered and disordered phases.

A similar trend is observed for the $d$-electron channel, with an additional reduction in loss within the WL phase. In the WL phase, the electron density is more uniform, providing a better match with the input configuration, which leads to a lower reconstruction error.

For each pair of $U$ and $T$, we collected four measures based on the autoencoders' loss functions: the mean losses and variances for both the $f$ and $d$ channels. Thus, each point in the phase diagram is represented by a four-dimensional vector, ${\bm a}_{U, T} = (m_d, m_f, v_d, v_f)$. We then applied the $k$-means clustering algorithm to group these vectors into $n_{\rm c}$ clusters. The value of $n_{\rm c}$, as a model hyperparameter, was set to be equal to or greater than the expected number of phases in the diagram. Figure~\ref{fig:aenc_loss}(c) shows the results of $k$-means clustering with $n_{\rm c}=4$, which provides a reasonable agreement with the phase transition boundaries. Increasing $n_{\rm c}$ further did not lead to any noticeable improvement in the delineation of phase boundaries.

\section{Conclusions \label{sec:Concl}}

To summarize, we showed that simple local PCA of $f$ and independently $d$ configurations can be used to automatically identify the phase boundary between the ordered and disorder phases, for both the first and second order phase transformation regimes, and that this procedure is straightforward to interpret. The difference between the EVRs of the $d$ and $f$ subsystems gives the position of the phase boundary between the regimes with weak and intermediate interactions in the disordered phase. This boundary coincides with the transition from the WL to the DI regime. This demonstrates that PCA can be used to investigate Anderson and weak localization in two-dimensional systems.  

The order-disorder phase boundary and the difference between the WL and DI regimes are robust and reflected in the $f$ and $d$ configurations. Consequently, other unsupervised techniques are also able to identify it, although with less precision than the local PCA. The prediction-based classification technique strongly depends on the predictor used and how the data are prepared. Yet, relatively simple neural networks, such as CNN and FCNN, are already sufficient to detect three distinct regimes. A standard k-means clustering applied to the mean values of the loss function calculated using the validation sample in the autoencoder approach highlights the OP, WL, and DI regimes as well. Nevertheless, the local PCA, which is the least computationally demanding and the simplest to implement of all methods presented here, offered the clearest and most precise identification of phase and regime boundaries.

What these methods have not been able to identify is the AI-MI phase boundary or equivalent boundaries in the ordered phase. The reason is that crucial information about the DOS gap on the Fermi level is not encoded in the $d$ and $f$ configurations. We conclude this on the basis that we have tested many other methods, which have not been presented here. They all failed in this task when the only input data were the configurations. For example, we tested several PCA kernels and some standard unsupervised methods for clustering without success. We used several other simple predictors. We also tested supervised learning, which, however, either conflated the AI-MI and WL-DI boundaries, or the supposed AI-MI boundary was very sensitive to the data chosen in learning.  Several of these techniques worked when we also included additional information in the training, e.g., the single-particle eigenenergy spectra or the actual value of the gap in the learning process. We do not present these results because they are trivial, and also the techniques used do not align with the goal of the paper of identifying phase boundaries just from the particle-occupancy snapshots.

Nevertheless, the inability to distinguish between gapped and gapless phases from configuration snapshots does not diminish the remarkable capability of the simple unsupervised methods presented here to automatically identify phases and regime boundaries in complex correlated systems. As formulated, these methods are readily applicable to any quantum-classical model of correlated electrons, including all variants of FKM~\cite{freericks2003exact} and Anderson model~\cite{evers2008anderson} as well as systems that combine classical spins with conducting electrons, which are relevant for various materials (e.g.,~\cite{liang2013nematic,buhler2000magnetic,yin2010unified}), spintronics (e.g.,~\cite{mondal2021when,smorka2022nonequilibrium}) or topology (e.g.,~\cite{hu2015interplay,diaz2021majorana,elbracht2020topological,maska2021unconventional}). Furthermore, since these methods operate on mean occupancies and have proven effective even for properties inherently quantum in nature, we anticipate their applicability to models extending beyond the single-particle paradigm.

\begin{acknowledgments}

This work was supported by the Ministry of Education, Youth and Sports of the Czech Republic through the e-INFRA CZ (ID:90254). L.F. and M.\v{Z}. acknowledge support from the Czech Science Foundation via Project No. 22-22419S. P.B. acknowledges the assistance provided by the  Operational Programme Johannes Amos Comenius of the Ministry of Education, Youth  and  Sport of the Czech Republic, within the frame of project Ferroic Multifunctionalities (FerrMion) [project No. CZ.02.01.01/00/22\_008/0004591], co-funded by the European Union. 
\end{acknowledgments}

%----------
\appendix

%-----
\section{Global PCA\label{app:globalPCA}}

\begin{center}
\begin{figure}[ht]
\includegraphics[width=1.0\columnwidth]{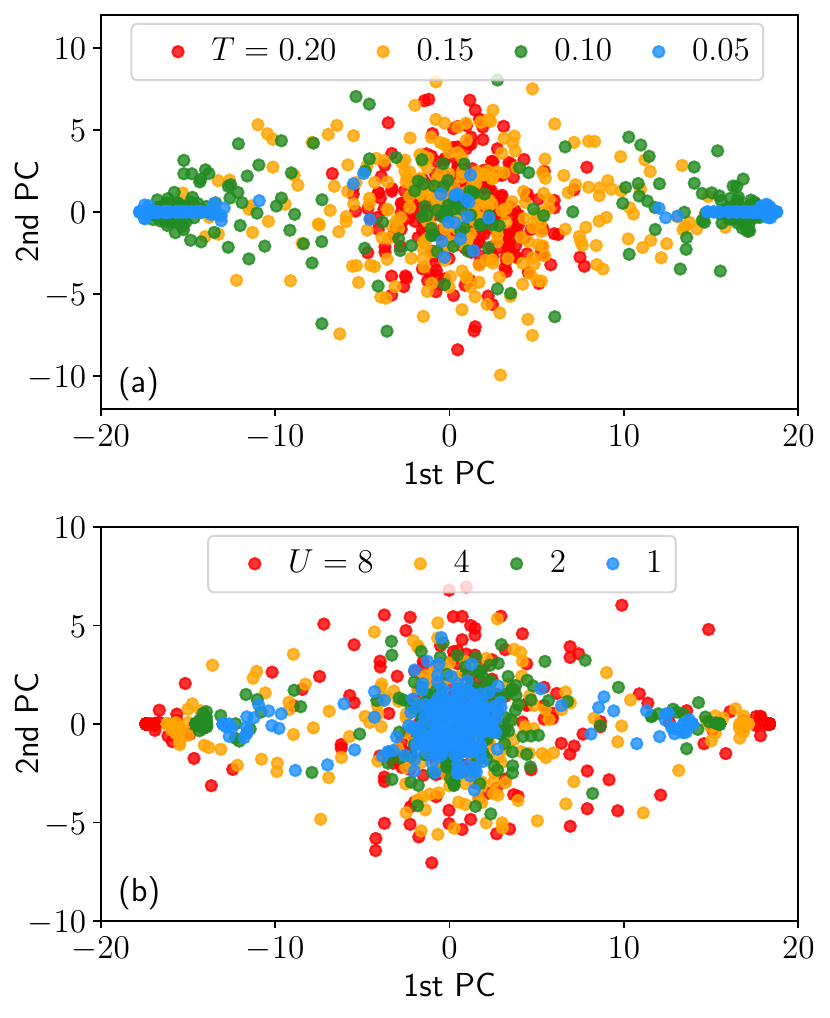}
\caption{A projection of combined $f$ and $d$ configurations into the two-dimensional plane spanned by the first two principal components obtained by the global PCA variant. (a) The different colors show the temperature at which the configurations have been obtained. (b)  The different colors show the coupling $U$ at which the configurations have been obtained.\label{fig:globalPCA}}
\end{figure}
\end{center}
The results presented in Sec.~\ref{sec:ResPCA} were obtained with the local variant of the PCA. Here we present some examples obtained with the more standard global variant.
To make the distinction clearer, we repeat that in the local approach
the PCA is applied independently to all combinations of $U$ and $T$ and for each of these combinations the EVR is calculated.
In the global approach, we fit a single global PCA model on the entire dataset, that is, for all $U$ and $T$ together. Then we transform the
data for each $U$-$T$ pair using this global PCA model.
Finally, the explained variance ratios for each $U$-$T$ pair are calculated based on this global transformation.
\begin{center}
\begin{figure}[ht]
\includegraphics[width=1.0\columnwidth]{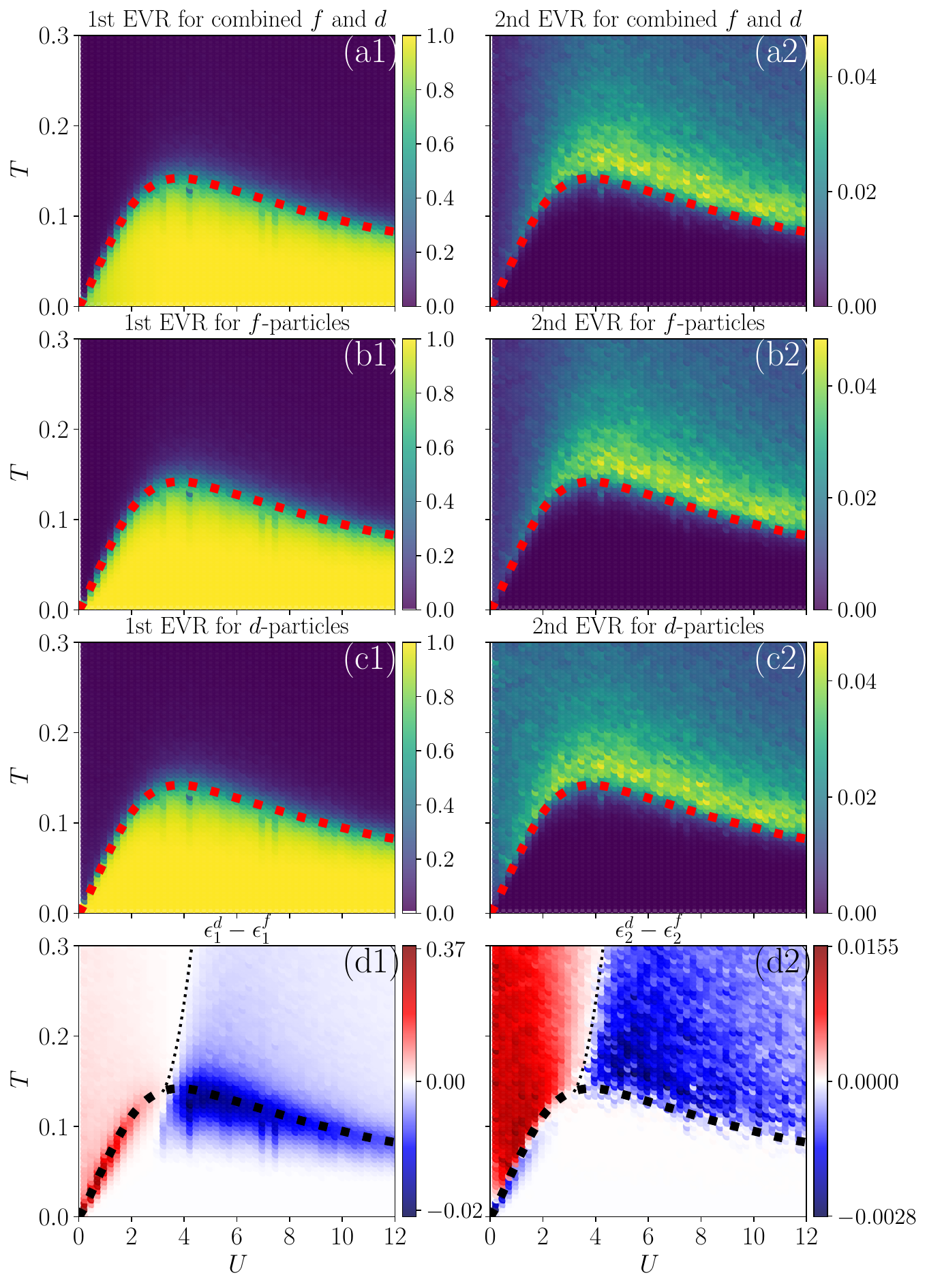}
\caption{ Results of the global PCA for $L=12\times 12$. Panel (a1) shows the value of the first EVR for the combined $f$ and $d$ particle configurations, panel (b1) for separated $f$-particle, panel (c1) separated $d$-particles and panel (d1) their difference.  Panel (a2) shows the value of the second EVR for the combined $f$ and $d$ particle configurations, panel (b2) for separated $f$-particle, panel (c2) separated $d$-particles and (d2), again, their difference.  The dashed lines signal the phase boundary between the ordered and disordered phase and between the WL and DI regimes respectively, taken graphically from ref.~\cite{antipov2016interaction}.  \label{fig:gPCAphd}}
\end{figure}
\end{center}
Figure~\ref{fig:globalPCA} shows a projection of PCA for combined $f$ and $d$ configurations into the two-dimensional plane spanned by the first two principal components for several temperatures. For clarity, only a fraction of the data is shown. Each chosen temperature [Fig.~\ref{fig:globalPCA}(a)] or interaction strength [Fig.~\ref{fig:globalPCA}(b)] is marked by points of different colors. In panel (a) at high temperatures, that is, above the critical one, the points concentrate in the center. At low temperatures a tendency to separate the data into two distinct regions is evident. However, the separation cannot be perfect because all the data for all $U$ are used and the critical temperature of the order-disorder transition is dependent on $U$. For the same reason, the separation according to $U$ shown in Fig.~\ref{fig:globalPCA}(b) is even less successful.

Figure~\ref{fig:gPCAphd} illustrates the explained variation ratio for the first and second components obtained from the global PCA analysis. Panels (a1) and (a2) depict the results for the combined $f$ and $d$ configurations, while panels (b1) and (b2) present the analysis for the 
$f$ configurations separately. Similarly, panels (c1) and (c2) show the results for the $d$ configurations, and finally, panels (d1) and (d2) display their differences.

The results demonstrate that the global PCA approach is capable of identifying the boundary between the ordered phase and the disordered one, see panels (a1)-(c2), and between the WL and DI regimes,  see panels (d1) and (d2). However, the transition is less sharp than for the local variant.
\begin{center}
\begin{figure}[ht]
\includegraphics[width=1.0\columnwidth]{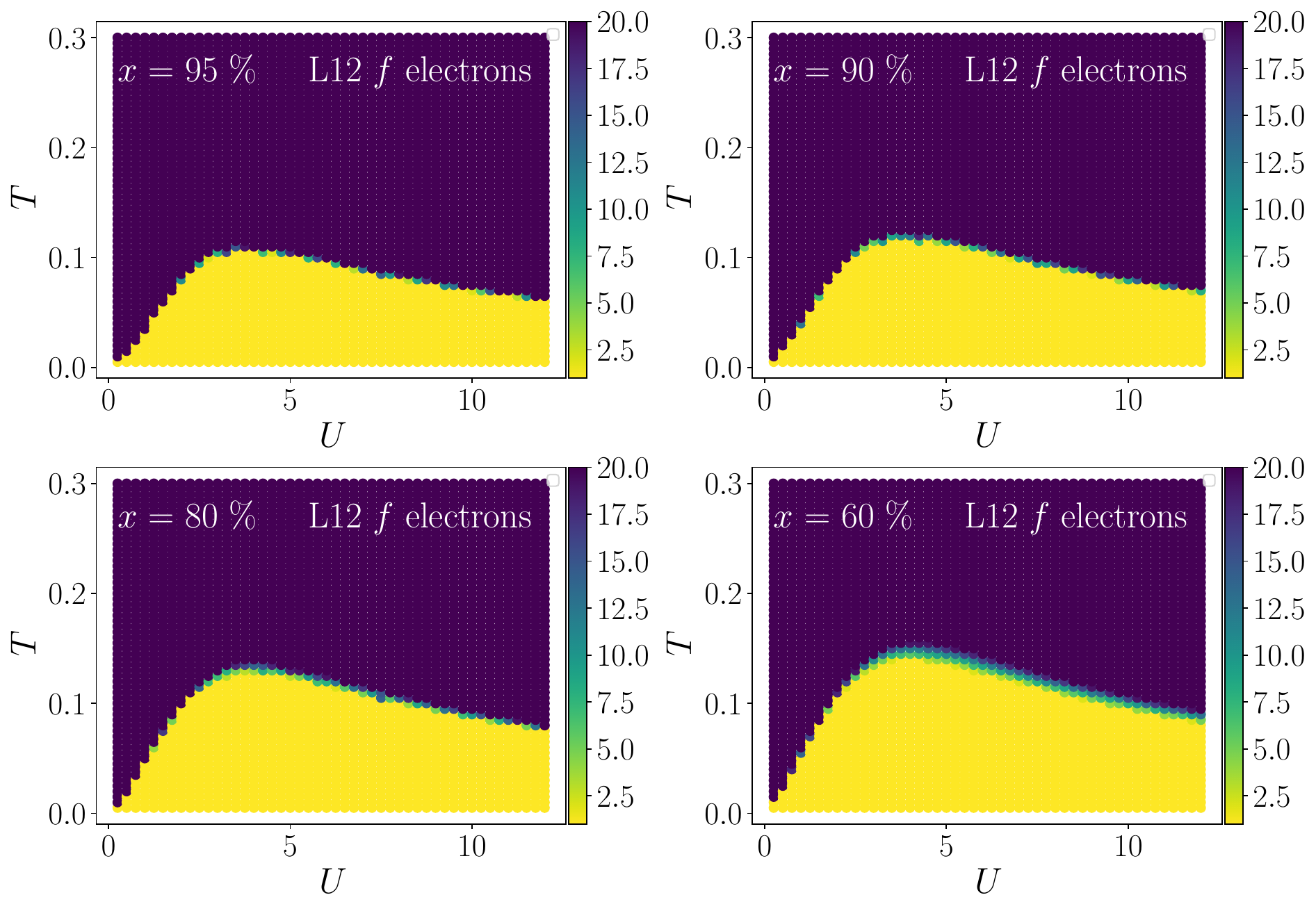}
\caption{Local PCA for $f$-particles and $L=12\times 12$ showing the number of principal components needed to explain $95\%$ (a), $90\%$ (b),  $80\%$ (c) and $60\%$ (d) of the total variance for each combination of analyzed parameters. Note that the colorbar was truncated at $N_\mathrm{PC}=20$ to better emphasize the transition area around the phase boundary.  \label{fig:localPCA_f}}
\end{figure}
\end{center}

\begin{center}
\begin{figure}[ht]
\includegraphics[width=1.0\columnwidth]{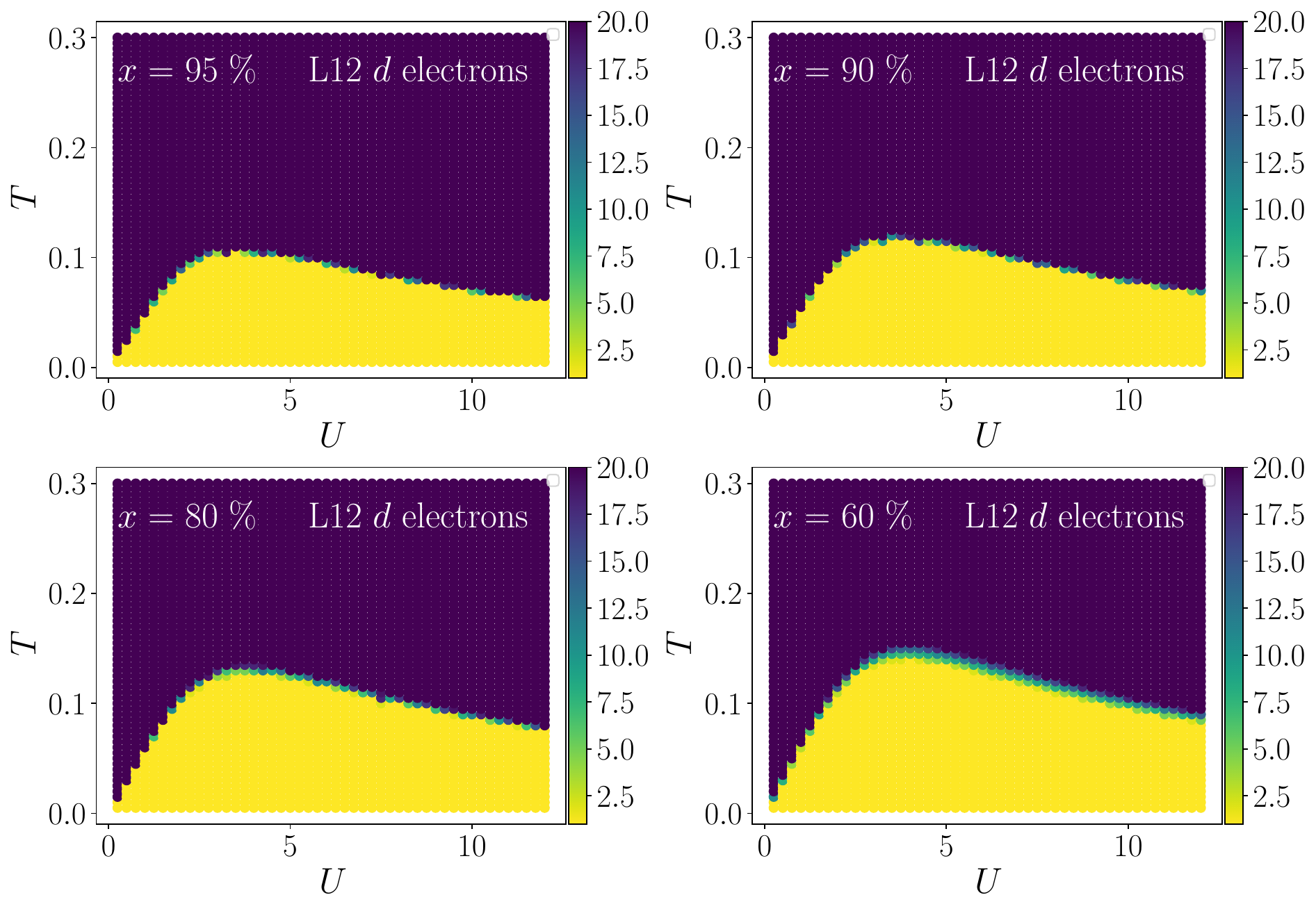}
\caption{Local PCA for $d$-particles and $L=12\times 12$ showing the number of principal components needed to explain $95\%$ (a), $90\%$ (b),  $80\%$ (c) and $60\%$ (d) of the total variance for each combination of analyzed parameters. Note that the colorbar was truncated at $N_\mathrm{PC}=20$ to better emphasize the transition area around the phase boundary.   \label{fig:localPCA_d}}
\end{figure}
\end{center}

%-----
\section{Local PCA\label{app:localPCA}}
Here we present some supporting results for the local PCA method. Figure~\ref{fig:localPCA_f} and Fig.~\ref{fig:localPCA_d}
show the analysis for $f$ and $d$ configurations and lattice size $L=12\times 12$, where the number of principal components needed to explain $95\%$ (a), $90\%$ (b),  $80\%$ (c) and $60\%$ (d) of the total variance is plotted using the color scheme. Note that the data are truncated at $N_\mathrm{PC}=20$ to better emphasize the transition area around the phase boundary. Still, the transition from $N_\mathrm{PC}=0$ to many is sharp, which justifies the usage of only the first two EVRs in our analysis.  
%---
%---
\section{Away From Half Filling \label{app:segregated}}
One can argue that PCA works well at a particle-hole symmetric point that enforces the half filling condition, because there the ground-state configuration forms a simple checkerboard structure. This configuration is particularly well-suited for linear transformations under PCA. Here, we briefly demonstrate that PCA can still be useful even far from this point, although the analysis becomes less straightforward.
\begin{figure}[htbp]
\includegraphics[width=1.0\columnwidth]{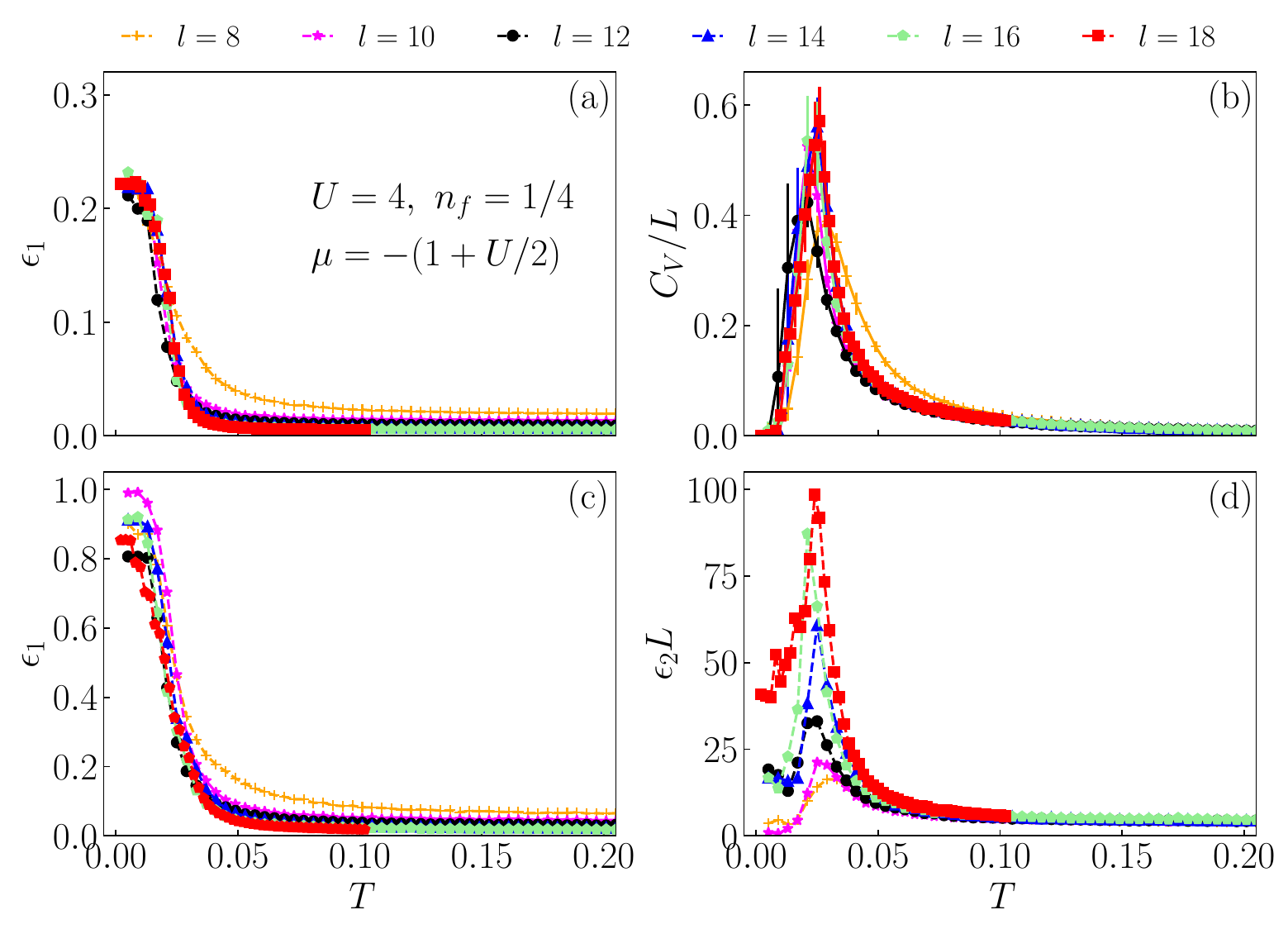}
\caption{Local PCA analysis for a case with $N_f$ fixed at $L/4$, $U=4$ and $\mu = -(1+U/2)$.  
(a) First EVR obtained from PCA applied directly to the MC $f$-configurations without preprocessing.  
(b) Approximate specific heat as a function of temperature.  
(c) First EVR computed for the structure factor of the $f$-configurations. 
(d) Second EVR computed for the structure factor of the $f$-configurations.\label{fig:EVRnoHFU4}}
\end{figure}

\begin{figure}[htbp]
\includegraphics[width=1.0\columnwidth]{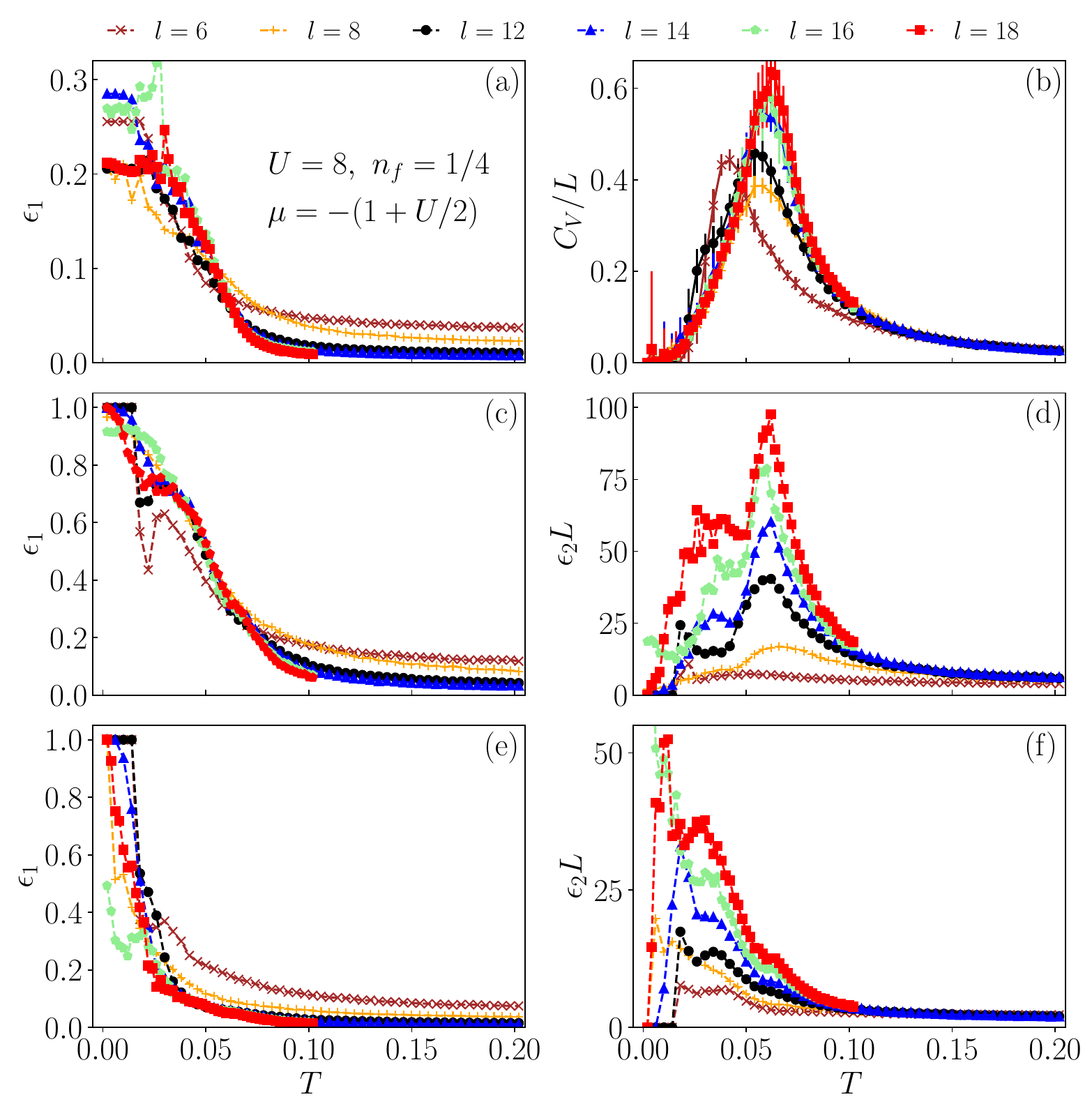}
\caption{Local PCA analysis for a case with $N_f$ fixed at $L/4$, $U=8$ and $\mu = -(1+U/2)$.  
(a) First EVR obtained from PCA applied directly to the MC $f$-configurations without preprocessing.  
(b) Approximate specific heat as a function of temperature.  
(c) First EVR computed for the structure factor of the $f$-configurations.  
(d) Second EVR computed for the structure factor of the $f$-configurations.  Panels (e) and (f) show first and second EVRs obtained for the $f$-configurations after removing symmetries following the procedure detailed in the text.\label{fig:EVRnoHFU8}}
\end{figure}

To this end, we investigate a case where $N_f$ is fixed at $L/4$ and the chemical potential is set to $\mu = -(1+U/2)$ (or equivalently, $\mu = -1$ in the absence of the half filling shift introduced in Eq.~\eqref{eq:Model}). This implies that we employ only canonical MC updates for the $f$-particles, while the number of $d$-particles is determined by $\mu$ and varies with temperature. We focus on the intermediate $U=4$ and the strong $U=8$ coupling regimes. The used $\mu$ sets $N_d/L$ between $0.22$ and $0.255$ for all lattice sizes considered, with its temperature dependence having a negligible effect on subsequent analysis.
\begin{figure}[htbp]
\includegraphics[width=1.0\columnwidth]{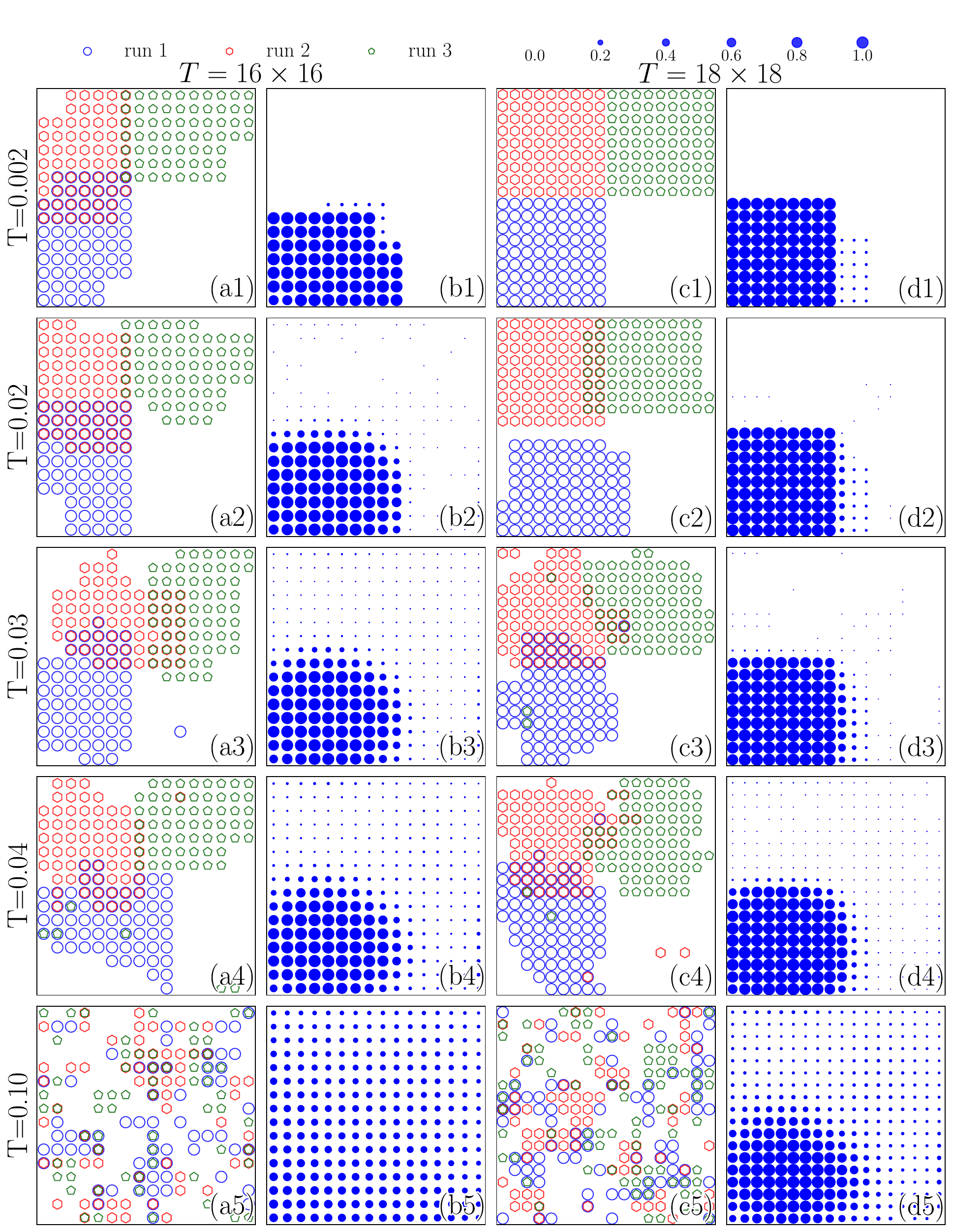}
\caption{Columns (a) and (c) show snapshots of three $f$-configurations from independent MC runs at various temperatures for $L=16\times 16$ (a) and $L=18\times 18$ (c). Columns (b) and (d) show mean $f$-particle occupancy after removing symmetries, following the procedure detailed in the text. The data were obtained for $U=8$ and $\mu=-(1+U/2)$.
\label{fig:snsh}}
\end{figure}

For these parameters, the ground state is expected to be a segregated phase (for similar regimes, see, e.g., ~\cite{lemanski2002stripe,zonda2012phase}), a result also confirmed by our calculations. This regime presents a greater challenge for PCA compared to the half filled case, as explained below.

For simplicity, we focus only on $f$-configurations. Figures~\ref{fig:EVRnoHFU4}(a) and ~\ref{fig:EVRnoHFU8}(a) display the first EVR as a function of temperature for different lattice sizes. It decreases from $\epsilon_1\approx 0.25$ to $1/L$ near the temperature corresponding to the maxima of the approximate specific heat,
\begin{equation}
C_L=\left(\langle E^2 \rangle - \langle E \rangle^2\right)/T^2,
\end{equation}
shown in Fig.~\ref{fig:EVRnoHFU4}(b), respectively in Fig.~\ref{fig:EVRnoHFU8}(b), where we neglect a small contribution from the fluctuations of $N_d$. This shows that already a direct EVR provides insight into the order-disorder transition. However, unlike the half filled case, $\epsilon_1$ does not approach one as $T\rightarrow 0$ and exhibits a much less smooth behavior. 

The former issue results in $\epsilon_2$ (not shown here) closely resembling $\epsilon_1$, thus lacking the characteristic peak observed in the half filling analysis. The fact that $\epsilon_1 \neq 1$ even at $T=0$ is a consequence of lattice symmetries, specifically, periodic translations, rotations, and mirror reflections of low-temperature configurations that are energetically equivalent. These symmetries are not fully captured by direct PCA. 

When applying PCA to the structure factor [Figs.~\ref{fig:EVRnoHFU4}(c),(d) and Figs.~\ref{fig:EVRnoHFU8}(c),(d)], which takes care of the translation symmetries, the first EVR approaches one for most lattices. This is due to the fact that the ground-state configuration is for most lattices a square cluster with the size $(L/2) \times (L/2)$ [see Fig.~\ref{fig:snsh}(c1)] or other regular shape. Therefore, other lattice symmetries either do not play a role or are taken care of by PCA. The reason why $\epsilon_1$ does not converge to one for all lattices is due to finite-size effects. Not all lattices have regular ground states [see, e.g., Fig.~\ref{fig:snsh}(a1)]. Various irregular, yet segregated, shapes have very similar energies, and the effective annealing process within our MC simulations does not always converge to the correct ground-state configuration. Nevertheless, even for these cases (e.g. $L=8\times 8$ and $L=16\times 16$ for $U=8$) the second EVR shows a clear peak aligned with the peak in $C_V$. 

What sets this case apart from the simpler CDW case is that both $\epsilon_1$ and $\epsilon_2$ exhibit an intermediate plateau-like feature between the $T \to 0$ limit and the transition temperature marking the onset of disorder. This behavior becomes evident when examining the evolution of $f$-configurations with temperature. 

Figure~\ref{fig:snsh} presents snapshots of $f$-configurations from three independent Monte Carlo runs at various temperatures (columns (a) and (c)). The corresponding panels in columns (b) and (d) display an approximation of mean occupancies. To construct these, we first applied all possible lattice symmetry transformations to each $f$-configuration, reshaped back into two dimensions, and retained the transformed configuration that minimized the value of  
\begin{equation}
d=\sqrt{\sum_{x,y} \left( f_{x,y}(x-d_x) \right)^2 + \left( f_{x,y}y \right)^2 },
\end{equation}
where $x$ and $y$ are the lattice coordinates, and $d_x$ is a small shift introduced to break the $x$-$y$ symmetry. 

Although this procedure tends to overestimate the degree of ordering, it effectively reveals the type of the order and illustrated the transition from ordered to disordered regime.

The melting of ordered configurations occurs in two stages. First, the low-temperature structures [e.g., the regular squares in panel (c1)] melt into general (irregular) island of connected $f$ particles [panels (c2)-(c4)] while the system remains in a segregated (ordered) regime. However, in PCA, these configurations are not equivalent under linear transformations, making this regime distinct and leading to transient features in the first and second EVRs.  

To demonstrate that these effects are not merely a consequence of lattice symmetries, we also applied PCA to the transformed configurations discussed earlier. The EVRs in Figs.~\ref{fig:EVRnoHFU8} (e) and ~\ref{fig:EVRnoHFU8}(f) clearly distinguish different stages of the melting process. This is also reflected, albeit to a lesser extent, in $C_V$, where the transition stage manifests itself as small humps on the left ridge of the peak. However, this feature in $C_V$ is sensitive to finite-size effects and requires more precise calculations for a detailed analysis.

Nonetheless, as shown here, PCA remains a valuable tool for analyzing the phase diagram even beyond the half filling case.

%\clearpage
%\pagebreak

%\bibliographystyle{apsrev4-2}
%\bibliography{fkm}


\end{document}